\documentclass[a4paper,11pt]{article}

\usepackage{jheppub} 

\usepackage[T1]{fontenc} 

\title{\boldmath Phase diagram of QCD chaos \\
in linear sigma models and holography}

\author{Tetsuya Akutagawa,}
\author{Koji Hashimoto,}
\author{Takeshi Miyazaki}
\author{and Toshihiro Ota}

\affiliation{Department of Physics, Osaka University, Toyonaka, Osaka 560-0043, Japan}

\emailAdd{akutagawa@het.phys.sci.osaka-u.ac.jp}
\emailAdd{koji@phys.sci.osaka-u.ac.jp}
\emailAdd{tmiyazaki@het.phys.sci.osaka-u.ac.jp}
\emailAdd{tota@het.phys.sci.osaka-u.ac.jp}

\abstract{
Measuring chaos of QCD-like theories is a challenge for formulating a novel characterization of 
quantum gauge theories.
We define a chaos phase diagram of QCD allowing us to locate chaos in the parameter space of 
energy of homogeneous meson condensates and the QCD parameters such as pion/quark mass.
We draw the chaos phase diagrams obtained in two ways: first, by using a linear sigma model, varying 
parameters of the potential, and second, by using the D4/D6 holographic QCD, varying the number of colors $N_c$ and the 't Hooft coupling constant $\lambda$.
A scaling law drastically simplifies our analyses, and we discovered that the chaos originates in the maximum of the potential, and larger $N_c$ or larger $\lambda$ diminishes the chaos. }

\preprint{OU-HET-966}

\begin{document} 
\maketitle
\flushbottom


\section{Introduction}

To complete the phase diagram of QCD is one of the goals of fundamental physics. Typically it is drawn with 
the axes of external parameters such as temperature and quark chemical potential.
The topological structure of the phase diagram, such as phase boundaries and order of phase transitions, 
depends largely on the constitution of the QCD Lagrangian. For example, 
the values of the quark masses and resultantly the number of flavors matter, and one could even change the number of colors $N_c$ and the coupling constant, relevant to the QCD scale.
While how the phase diagram changes as these parameters of QCD vary has been explored in various cases,
the reason why our universe is made of the QCD having such a phase diagram has not been understood.

Characterization of quantum field theories is not just by parameters in Lagrangians. In this paper we adopt 
{\it chaos} as a dynamical characterization of quantum field theories. Chaos could be an index to recognize how
complicated a given quantum field theory is. From the lessons of the standard QCD phase diagrams,
we better explore a possible ``phase diagram of QCD chaos.'' Our goal of this paper is to provide a definition
of a phase diagram of chaos in QCD-like gauge theories, and to draw it concretely by using some approximation
of QCD or effective models of QCD. The chaos could add an ``dynamical'' axis to the QCD phase diagram
and extend the whole structure,
and might characterize existent QCD phases from a higher-dimensional perspective.

For defining a phase diagram of QCD chaos, a major obstacle is the quantum nature of QCD. In fact, 
chaos can be clearly defined as behavior of a classical motion of variables in dynamical systems. 
The QCD observables which are responsible for the standard phase diagram are purely quantum, as is obvious 
with the example of the chiral condensate. Therefore, we need a ``classical'' limit maintaining 
the quantum properties of QCD. This seem-to-contradicting limit is achieved once we look at low energy effective field theory of QCD. Linear sigma models, capturing mainly the global symmetry structure of QCD, are classically 
written by mesons which are quantum bound states of quarks. In \cite{Hashimoto}, a chaos was discovered 
in the linear sigma model.
In the first part of this paper, we employ the linear sigma model with a single flavor for drawing the
phase diagram of QCD chaos.

%
%

A deficit of the effective models is that the parameters of the model depend on the QCD parameters only implicitly. For example, the potential of the linear sigma models should depend on quark masses and QCD scales,
but the dependence is known only when one solves QCD explicitly. 
A solution to this problem is to infer top-down holographic QCD models \cite{Karch,Kruczenski,Sakai1}
based on the AdS/CFT correspondence \cite{Maldacena}. For large $N_c$ and at strong coupling,
QCD-like gauge theories have a dual classical description, where a dynamical potential for mesons
can be explicitly deduced as a function of the original QCD parameters. In the second part of this paper,
we shall use a holographic QCD model to draw the phase diagram of QCD chaos.
The holographic model we employ is the single-flavor 
D4/D6 system \cite{Kruczenski}, because it describes a non-supersymmetric QCD and shares a symmetry breaking pattern with the linear sigma model.

There exists another virtue of employing the AdS/CFT for the chaos analyses.
Currently, it is known that quantum chaos plays a crucial role to characterize black holes in the AdS/CFT \cite{Maldacena2,Sekino}, and a maximal chaos is thought of as a criterion for 
a QFT to have a gravity dual \cite{Sekino,Shenker1,Shenker2,Maldacena3}. 
The chaos index is a Lyapunov exponent defined through out-of-time-orderd correlators\cite{Shenker1,Shenker2,Maldacena3,Larkin,Kitaev,Sachdev,Kitaev2} in QFT, and corresponding
exponent should be observed in chaotic
motion of the fundamental string or D-branes in a curved spacetime \cite{Zayas,Basu1,Basu2,Asano,Ishii,Tanahashi1,Tanahashi2}. 
A holographic study of chaos of meson condensate was provided in \cite{Hashimoto} for the 
supersymmetric D3/D7 model \cite{Karch,Kruczenski2}.
For QCD-like theories, to explore thoroughly the chaos phase diagrams in their gravity duals 
will help us finding directions to uncover the mystery of emergent spacetime.

The phase diagram of QCD chaos we draw, using the linear sigma model and the D4/D6 holographic model, is a plane spanned by
the axis of the quark mass and that of the energy density. The plane is divided into regions with/without chaos for the homogeneous
motion of the sigma meson and the pi meson. A short summary of the characteristic features of the phase diagram which we obtain is
as follows:
\begin{itemize}
\item
For a fixed quark mass, the chaos appears only for a middle range of the energy density. At low energy, or at high energy, there is no chaos.
\item
When the quark mass vanishes, chaos disappears. Turning on the quark mass lets the chaos region in energy grow.  
\item
The energy scale of the chaos is centered at the local maximum of the linear sigma model potential, not the saddle point.
\item
The system is less chaotic for larger 't Hooft coupling $\lambda$ and larger $N_{c}$, since the center of the chaos region increases.
\end{itemize}
For our result of the phase diagrams, see figure \ref{sigmaphasediagram1} for the linear sigma model 
and figure \ref{fig:diagramrinfty} for the D4/D6 model.
To draw the chaos phase diagram, we develop technical tools such as scaling symmetries of the linear sigma model
and the holographic model, and a map between the two models.

%
%

This paper is organized as follows: 
In section \ref{sec:linearsigma}, we study the chaos of a linear sigma model, and 
draw a phase diagram of chaos for the quark mass and the energy density. 
We determine the topology of the chaos phase in the diagram, and locate the origin of the chaos.
In section \ref{sec:D4D6}, by using the holographic D4/D6 model and comparing the system with the linear sigma model, 
parameters in the linear sigma model are expressed as a function of gauge theory parameters such as $N_c$ and $\lambda$. 
We study the parameter dependence of the chaos of mesons.
Section \ref{sec:conclusion} is dedicated to conclusions and discussions on implications of our chaos phase diagram, with future directions.

\section{Phase diagram of chaos in a linear sigma model}
\label{sec:linearsigma}

In this section, we draw a phase diagram of chaos in the linear sigma model for a single flavor QCD at low energy, ignoring axial anomaly. 
The phase diagram locates the region where the 
chaos is found in the motion of homogeneous configurations of the sigma meson and the pion, in the plane spanned by
the pion/quark mass and the total energy density of the mesons.
To find the chaos, we use Poincar\'e sections. 
The topology of the phase diagram is found to detect the symmetry restoration,
and suggests the origin of the chaos, which is discussed to be consistent with the standard phase diagram of finite temperature QCD.

%
%
%

\subsection{Review of a linear sigma model and its chaos}

In this subsection, we review the linear sigma model and its chaos found in \cite{Hashimoto}. 
The most popular low energy effective action for the chiral condensate of QCD is the linear sigma model. 
It describes a universal class of theories governed by the chiral symmetry via the spontaneous and the explicit breaking. 
The linear sigma model is known to have chaos \cite{Hashimoto} in the motion of the homogeneous meson condensates. 
In this paper we focus on the phase diagram of the chaos by looking at the parameter dependence of the chaos.

For the preparation for later subsections, here we review the linear sigma model.
The action has a chiral U(1)$_A$ symmetry with an explicit breaking term:
\begin{align}
S &= \int d^4 x\left\{-\frac{1}{2}[(\partial_\mu \sigma)^2+(\partial_\mu \pi)^2] -V\right\} \ , \\
V &:= \frac{\mu^2}{2}(\sigma^2+\pi^2)+\frac{g_4}{4}(\sigma^2+\pi^2)^2+a\sigma+V_0 \ .
\end{align}
For simplicity, we consider only a single flavor case and ignore the axial anomaly. $\sigma (x^{\mu})$ and $\pi(x^{\mu})$ are fields whose fluctuations provide a sigma meson field with the mass $m_\sigma$ and a neutral pion field with the mass $m_{\pi}$, respectively. $\mu^2, g_4, a$ are parameters and a constant $V_0$ is just for shifting the vacuum energy to zero.

\begin{figure}[tbp]
\centering
\includegraphics[width=80mm,bb=0 0 1024 768,trim=100 0 0 100]{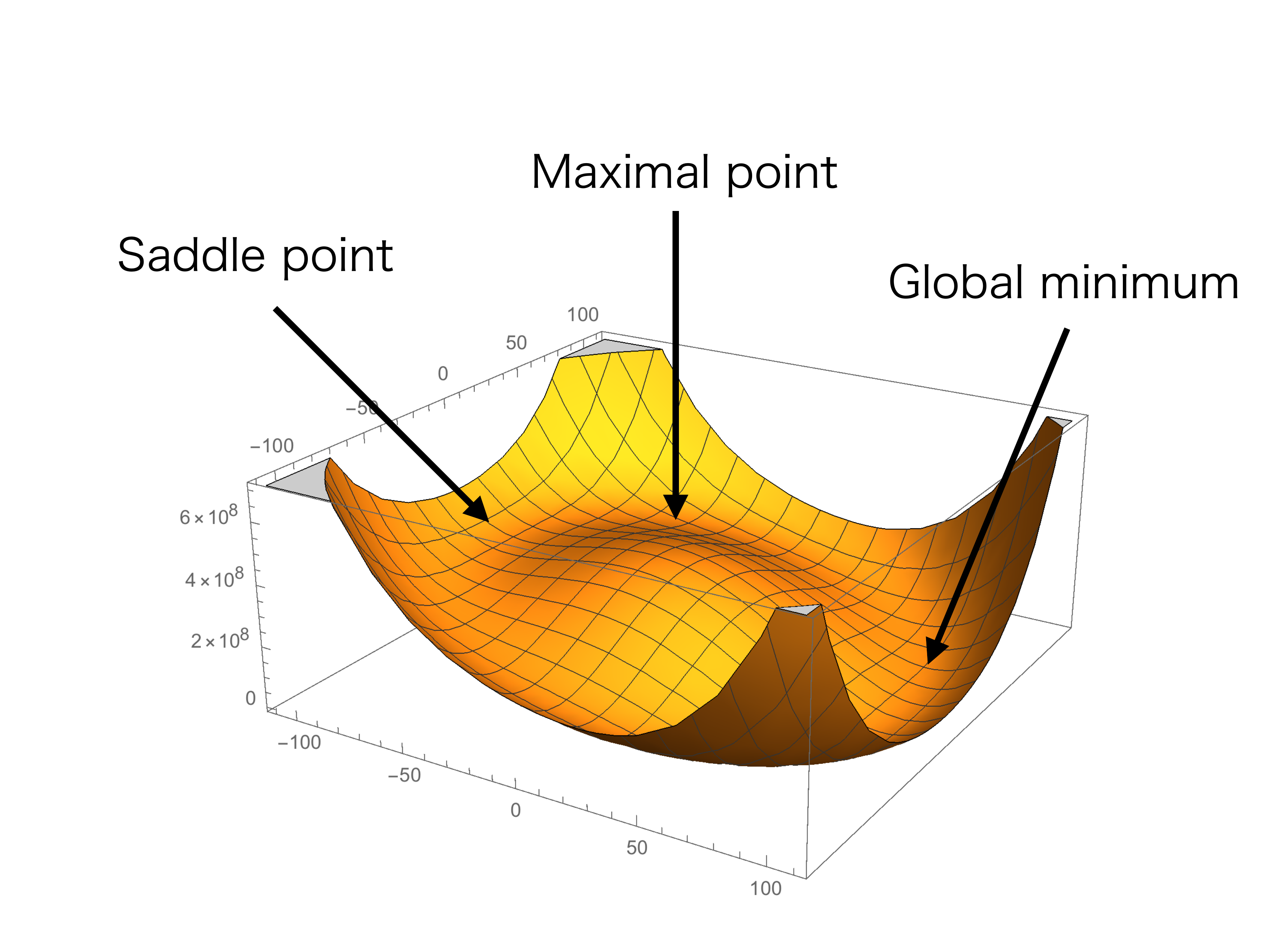}
\caption{The potential of the linear sigma model. The horizontal axes are $\sigma$ and $\pi$. The global minimum is at $(\langle \sigma \rangle,\langle\pi \rangle)=(f_{\pi},0)$. }
\label{spotential}
\end{figure}

To extract chaos from the motion of the meson condensates, the simplest assumption is to 
consider spatially homogeneous fields $\sigma (t), \pi(t)$. Then, the Hamiltonian becomes
\begin{align}
H=\frac{1}{2}(p_{\sigma} ^2+p_{\pi}^2)+\frac{\mu^2}{2}(\sigma^2+\pi^2)+\frac{g_4}{4}(\sigma^2+\pi^2)^2+a\sigma+V_0 \ ,
\end{align}
and the potential of the model is shown in figure \ref{spotential}. 
A static solution to the equation of motion given by this Hamiltonian is $(\sigma,\pi)=(f_{\pi},0)$, and $f_{\pi}$ satisfies
\begin{align}
\mu^2 f_{\pi}+g_4 f_{\pi} ^3 +a=0 \ .
\label{eq:fpi}
\end{align}

The model has three physical parameters: $\mu^2, g_4$ and $a$.
These parameters of the linear sigma model have the following relations to observed quantities: 
\begin{align}
2\mu^2=-m_{\sigma} ^2 +3m_{\pi} ^2 \ , \quad
g_4=(m_{\sigma} ^2-m_{\pi}^2)/(2f_{\pi} ^2) \ , \quad
 a=-m_{\pi} ^2f_{\pi} \ . 
\end{align}
Obviously, the parameter $a$ describes the explicit breaking of the axial symmetry, and represents the quark mass.
When we choose experimental values
$m_{\sigma}=500$ MeV, $m_{\pi}=135$ MeV, and $f_{\pi}=93$ MeV,  
the parameters in the linear sigma model are found as 
$\mu^2=-9.77 \times 10^{4}$ MeV$^2$, $g_4=13.4$, $a=-1.70 \times 10^6$ MeV$^3$. We hereafter call these values as experimental values.

\begin{figure}[tbp]
\begin{minipage}{0.5\hsize}
  \centering
  \includegraphics[width=25mm,bb=0 0 415 256,trim=150 -80 0 0]{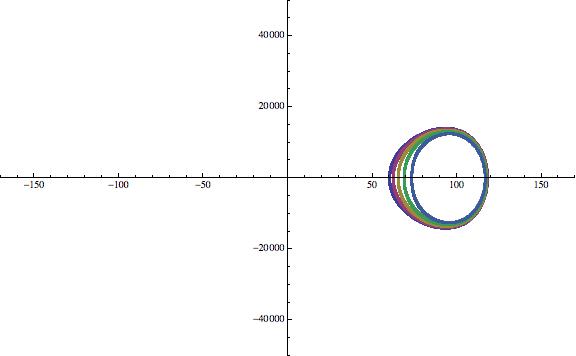}
\end{minipage}
\begin{minipage}{0.5\hsize}
  \centering
  \includegraphics[width=25mm,bb=0 0 415 256,trim=150 -80 0 0]{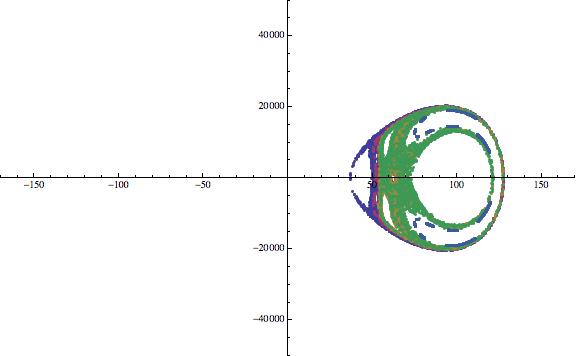}
\end{minipage}\\
\begin{minipage}{0.5\hsize}
  \centering
  \includegraphics[width=25mm,bb=0 0 415 256,trim=150 0 0 -50]{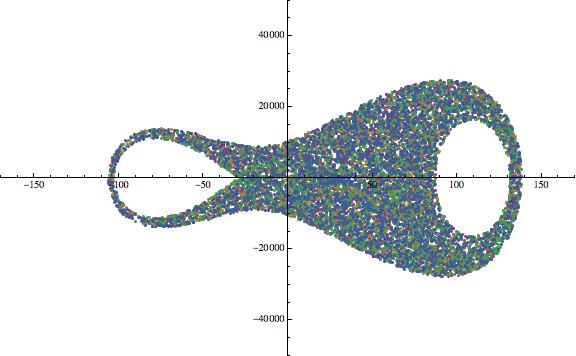}
\end{minipage}
\begin{minipage}{0.5\hsize}
  \centering
  \includegraphics[width=25mm,bb=0 0 415 256,trim=150 0 0 -50]{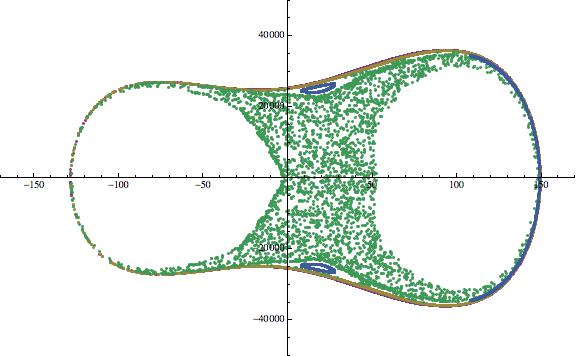}
\end{minipage}
\caption{The Poincar\'e sections in the linear sigma model (color online). The horizontal axis is $\sigma$, while the vertical axis is $p_{\sigma}$. The section is chosen as $\pi=0$. The energies are chosen as $E^{1/4}=100, 120, 140$ and 160 MeV in the top-left, top-right, lower-left and lower-right figures, respectively. A further increase of 
the energy results in just regular orbits.} 
\label{sigmaPS}
\end{figure}

At the experimental values, the Poincar\'e sections for the homogeneous motion of the mesons are plotted in figure \ref{sigmaPS}.
At the middle range of the energy, the Poincar\'e section contains a region made of scattered points, which shows the chaos.
This chaos at the experimental values of the sigma model parameters was found in \cite{Hashimoto}.

\subsection{Phase diagram of chaos}

Let us proceed to draw the phase diagram of chaos for the linear sigma model. First of all, the model has the three parameters $\mu^2, g_4, a$. 
Chaos depends on the energy, so the phase diagram is drawn in a four dimensional space spanned by $\mu^2, g_4, a$ and the energy.
There is a way to reduce the dimensionality of the phase diagram. We find that 
the model is invariant under the following scaling transformation: 
$\sigma \to \alpha \sigma,\ \pi \to \alpha \pi,\ t \to \beta t,\ \mu^2 \to (1/\beta^2)\mu^2,\ g_4 \to (1/\alpha^2 \beta^2)g_4,\ a \to (\alpha/\beta^2)a$,\ and $H \to (\alpha^2 /\beta^2)H$. 
Two parameters of the linear sigma model can be fixed by the degrees of freedom of the scaling transformations.  Therefore, 
it is enough to study Poincar\'e sections  with varying only one parameter among $\mu^2, g_4, a$, and the phase diagram
is a two-dimensional plot, which is much easier to understand.

In this paper we choose $a$ as the representative parameter of the system
because we are particularly interested in the relation between explicit symmetry breaking and chaos.
In fact, for the case of $a=0$, the linear sigma model is integrable, hence chaos should not appear. 
This fact is a good starting point for our technical analyses and also for a physical understanding of the relation
between the chaos in the linear sigma model and the quark mass.
In addition, due to 
a ${\mathbf Z}_2$ symmetry of the system, $\sigma \to -\sigma$ and $a\to -a$, the sign of $a$ does not affect the equations of motion. 
Hence in our analyses we vary $|a|$ and fix $\mu^2=-9.77 \times 10^{4}$ MeV$^2$ and $g_4=13.4$.
At the latter values, the choice $a=-1.70 \times 10^6$ MeV$^3$ brings us back to
the experimental values.

\begin{figure}[tbp]
\centering
\includegraphics[width=90mm,bb=0 0 1024 768,trim=120 90 50 0]{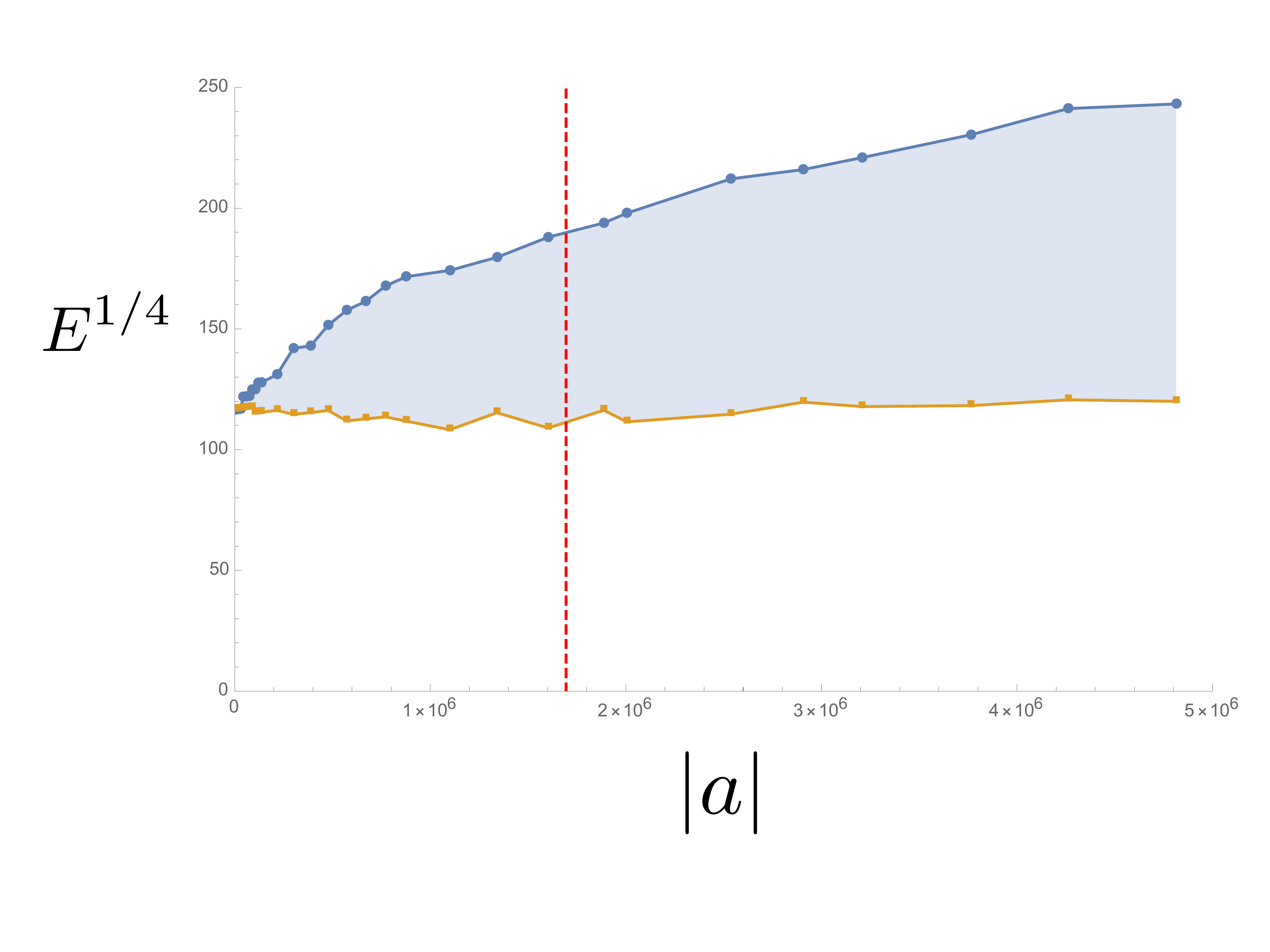}
\caption{The phase diagram of chaos in the linear sigma model. In the shaded region, the system exhibits chaos. At each energy, initial conditions 
for the meson motion are chosen as $(\pi,\, p_{\sigma},\, p_{\pi}) = (0.05 f_{\pi} \times i,\, 0,\, 0)$, $i = 1,\dots,7$, and $(\sigma,\, p_{\sigma},\, p_{\pi}) = (0.05 f_{\pi} \times j,\, 0,\, 0)$, $j = \pm 1,\dots,\pm 10$. The dashed line corresponds to the experimental values, {\it i.e.} the realistic QCD. }
\label{sigmaphasediagram1}
\end{figure}

By drawing numerically 
the Poincar\'e sections at each energy and each $|a|$
and by checking whether the section has a scattered plot or not, we 
obtain a phase diagram of the chaos in the linear sigma model. 
The resultant phase diagram is shown in figure \ref{sigmaphasediagram1}. 
The dashed line of $|a|=1.70 \times 10^6$ MeV$^3$ in figure \ref{sigmaphasediagram1} corresponds to the experimental values. 

We find that the phase diagram has an extremely simple structure. At $|a|=0$, there is no chaos, which is consistent with the integrability. Then increasing $|a|$ allows chaos in a small region of energy density 
around $E^{1/4}\sim 120$MeV. The chaos region gradually expands for larger $|a|$.
At the experimental value $|a|=1.70 \times 10^6$ MeV$^3$, the chaos region is
found as $110$MeV $< E^{1/4}< 190$MeV, which is consistent with what was found in \cite{Hashimoto}. 
Since $|a|$ is the parameter that breaks U(1)$_A$ symmetry, we can interpret that the chaotic region become wider when the symmetry breaking term contributes very much. 

\begin{figure}[tbp]
\centering
\includegraphics[width=90mm,bb=0 0 1024 768,trim=130 80 40 0]{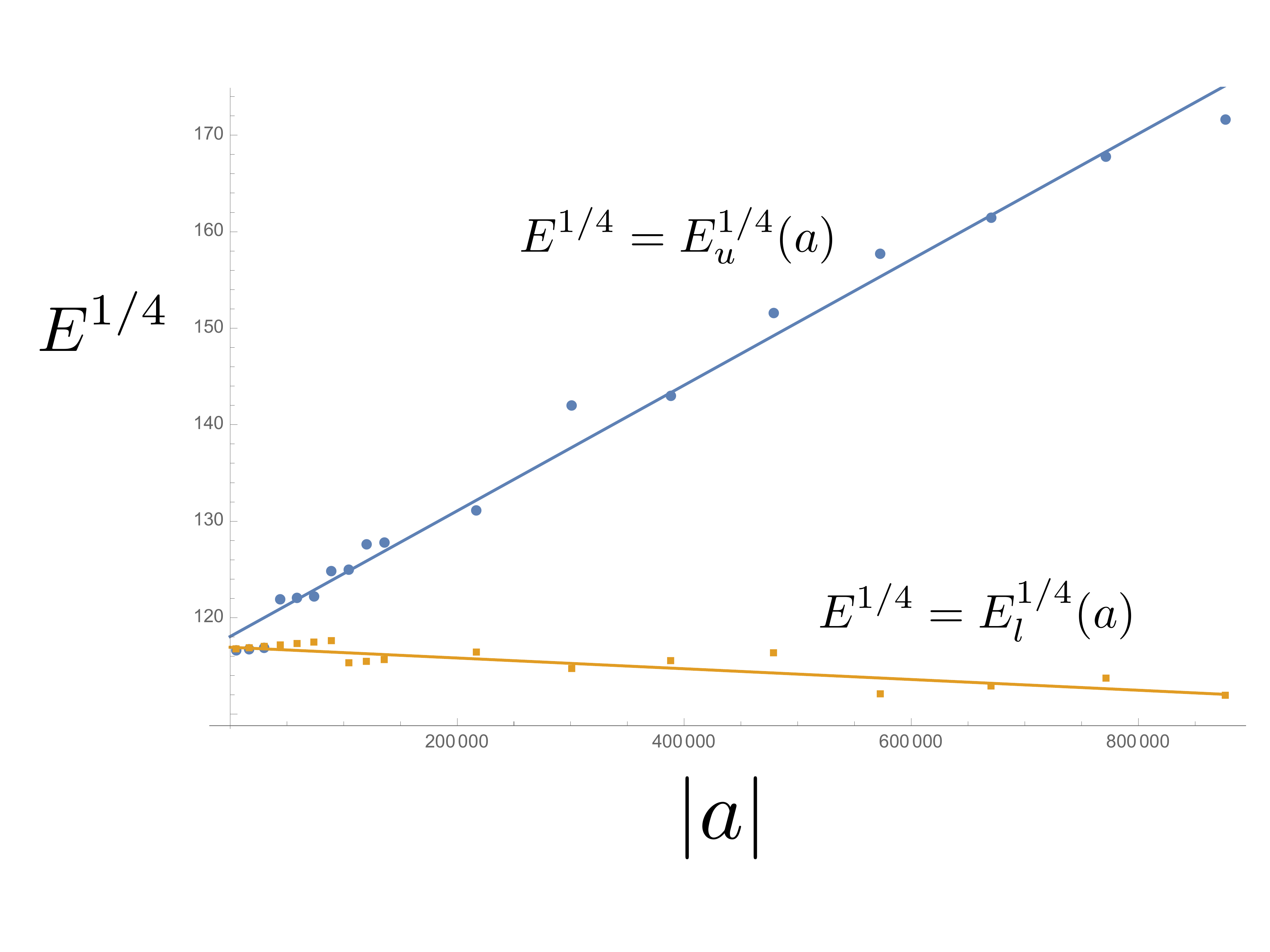}
\caption{The fitting functions for minimum and maximum values between which energies chaos exists. We calculate the fitting functions in the region $|a|\leq 8.77 \times 10^{5}$ MeV$^3$.}
\label{sigmaphasediagram2}
\end{figure}

The phase diagram of figure \ref{sigmaphasediagram1} suggests that for any nonzero infinitesimal $|a|$
a chaos region exists. To check this statement, we fit the chaos phase boundary by straight 
lines around $|a|=0$.
See figure \ref{sigmaphasediagram2}. 
We use linear regression to fit the phase boundaries in the range $|a|\leq 8.77 \times 10^{5}$ MeV$^3$. 
The obtained fitting function for the upper boundary of the chaos region is 
\begin{align}
E_u ^{1/4}(a)=1.18 \times 10^{2}+6.51 \times 10^{-5} \, |a| \ ,
\end{align}
and the standard deviation for the intercept is $2.07$ MeV. The fitting function for the lower boundary is 
\begin{align}
E_l ^{1/4}(a)=1.17 \times 10^{2}-5.58 \times 10^{-6} \, |a| \ ,
\end{align}
and the standard deviation is $9.42 \times 10^{-1}$ MeV. 
The two functions cross at $a=0$, within the numerical error bars. Therefore, it is consistent to interpret 
figure \ref{sigmaphasediagram1} as having chaos even for infinitesimal $|a|$.

\subsection{The origin of chaos}

Why does the chaos region look like emanating from a certain energy density at $|a|=0$?
The physical origin of the entire region of the chaos should come from the meaning of this energy density
$E^{1/4}\sim 120$ MeV at $|a|=0$.

\begin{figure}[tbp]
\centering
\includegraphics[width=90mm,bb=0 0 1024 768,trim=120 60 70 0]{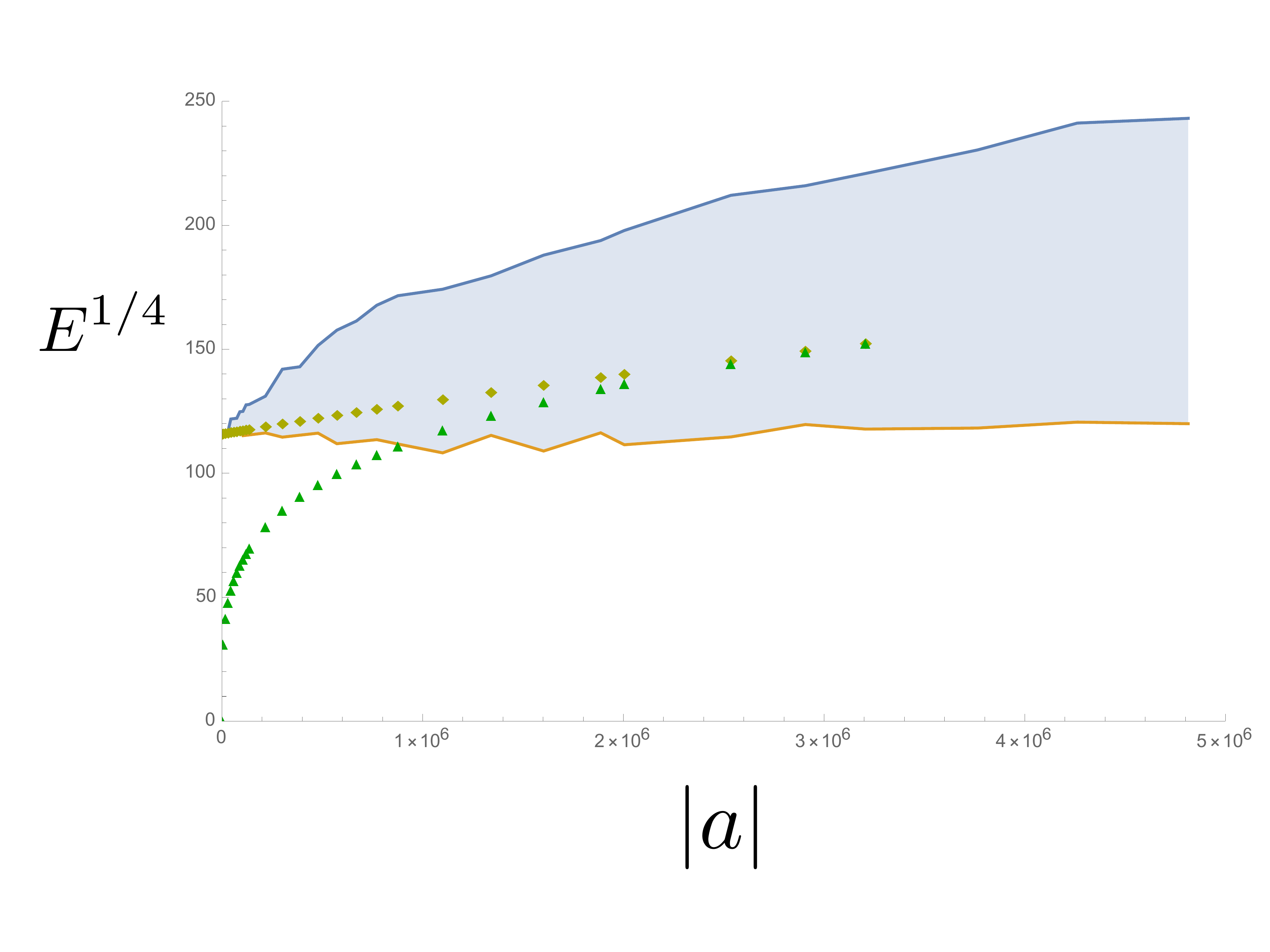}
\caption{The energy of chaos region. $\Diamond$ and $\triangle$ denote the maximal point and the saddle point in the potential, respectively. The maximal point and the saddle point in the potential  do not exist when $|a| >3.76 \times 10^6 $ MeV$^{3}$.}
\label{sigmaphasediagram3}
\end{figure}

The typical scale of the sigma model is determined by the potential, in which there exist a saddle point and
the potential maximum, see figure \ref{spotential}. 
In figure \ref{sigmaphasediagram3}, we show the energies of the maximal point and the saddle point 
in the potential, in the phase diagram.
From this figure it is obvious that the chaos is from the maximal point of the potential, not from the saddle point.
In general, it is known that chaos is associated with saddle points in a potential. However in our linear sigma model, the chaos is associated with the maximal point rather than the saddle point.

%
%

Although technically we are not sure why the saddle point of the potential of the linear sigma model 
does not give chaos, the behavior of having chaos around the maximum of the potential is consistent with
the thermal phase transition in QCD.
The classical chaos has a positive Lyapunov exponent, which is interpreted as Kolmogorov-Sinai entropy rate
through Pesin's relation. On the other hand, in thermal phase transitions entropy is produced. So
assuming that the information-theoretic entropy is regarded somehow as a statistical entropy\footnote{
See for example \cite{Sasa}.},
we find it natural to have the chaos at the maximum of the potential, because 
the maximum point of the linear sigma model
potential is generally related to the thermal
phase transition. Note that the saddle point is nothing related to the thermal phase transition. Even at $|a|\to 0$
when the saddle energy goes to zero, the thermal phase transition is present at a non-zero finite energy scale (temperature).

In this section, we have explored chaos in the linear sigma model and found the phase diagram, figure
\ref{sigmaphasediagram1}. The diagram was written in the space of $|a|$, one of the parameters in the potential
of the linear sigma model. 
%
Yet the explicit dependence of QCD parameters such as the number of colors $N_c$ and the 't Hooft coupling
constant $\lambda$ is hidden.
In the next section, we compare the action of the linear sigma model with the one of the D4/D6 system 
and find a relation of parameters between the two models.

\section{Phase diagram of chaos in D4/D6 holographic QCD}
\label{sec:D4D6}

In the previous section, we find that the chaos of the linear sigma model varies with a parameter $a$, {\it i.e.} 
quark mass, or pion mass. 
In this section, we draw a phase diagram in a plane spanned by QCD parameters such as quark mass, 't Hooft 
coupling constant $\lambda$ and the number of colors $N_c$. For that purpose, we need to find relations between
the linear sigma model parameters and the QCD parameters, that is effectively telling that we have to solve QCD
at low energy. We use the AdS/CFT correspondence \cite{Maldacena}, in particular the D4/D6 holographic QCD model which suits our purpose: it gives a gravity dual of a non-supersymmetric large $N_c$ QCD with a single flavor. The symmetry structure is the same as that of the linear sigma model, and one can introduce a quark mass
to that model.

To obtain the phase diagram of chaos, we first construct a map between the linear sigma model and
the D4/D6 model. Comparing the effective actions of the low energy fluctuations of the models
provide explicit relations between the models. Then using the relations, we recast our results in section \ref{sec:linearsigma} to the D4/D6 model.
We will see that the phase diagram 
shows that the system is less chaotic for larger 't Hooft coupling $\lambda$ and larger $N_{c}$.

\subsection{Review of the D4/D6 system}
\label{sec:d4d6}

Let us first give a brief review of the D4/D6 system \cite{Kruczenski}, to fix our notations and
to prepare for the study in the later subsections.

The D4 background we consider in what follows consists of $N_{c}$ D4-branes with one of the spatial world-volume directions wrapping on $S^{1}$, along which anti-periodic boundary conditions are imposed on fermions. 
This background corresponds to the holographic dual of a four dimensional pure Yang-Mills theory at low energy \cite{Witten}. 
The supergravity solution for the $N_{c}$ D4-branes in such a configuration takes the form
\begin{align}
ds^{2} &= \left( \frac{U}{R} \right)^{3/2}\left( \eta_{\mu\nu}dx^{\mu}dx^{\nu} + f(U)d\tau^{2} \right) + \left( \frac{R}{U} \right)^{3/2}\frac{dU^{2}}{f(U)} + R^{3/2}U^{1/2}d\Omega_{4}^{2} \label{eq:d4soliton} \ , \\
 e^{\phi} &= g_{s}\left( \frac{U}{R} \right)^{3/4} \ , \quad F_{4}=\frac{2\pi N_{c}}{V_{4}}\epsilon_{4} \ , \quad f(U)=1-\frac{U_{\text{KK}}^{3}}{U^{3}} \ .
\end{align}
\begin{table}[tbp]
\caption{The configuration of $N_{c}$ D4-branes and a single D6-brane. }
\begin{center}
\begin{tabular}{|c||c|c|c|c|c|c|c|c|c|c|}\hline
   & 0 & 1 & 2 & 3 & 4 & 5 & 6 & 7 & 8 & 9 \\ \hline
 $N_{c}$ D4 & \checkmark & \checkmark & \checkmark & \checkmark & \checkmark &  &  &  &  &  \\ \hline
         D6 & \checkmark & \checkmark & \checkmark & \checkmark &  & \checkmark & \checkmark & \checkmark &  &  \\ \hline
 \end{tabular}
 \label{tab:d4d6}
 \end{center}
\end{table}

\noindent
We are following the notations used in \cite{Kruczenski}. 
The coordinates $x^{\mu}$ and $\tau$ are the directions along the D4-branes, and the $\tau$ direction is compactified. 
$d\Omega_{4}^{2}$ and $\epsilon_{4}$ are the line element and the volume form on a unit $S^{4}$, respectively, and $V_{4} = 8\pi^{2}/3$ is its volume. 
$R$ is given by the string length $l_{s}$ and the string coupling $g_{s}$ as $R^{3} = \pi g_{s} N_{c} l_{s}^{3}$, and $U_{\text{KK}}$ is a constant parameter. 
The coordinate $U$ is a radial direction transverse to the D4-branes, and is bounded from below by the condition $U \geq U_{\text{KK}}$. Note that there is no spacetime in the region $U<U_{\text{KK}}$.
To avoid a conical singularity at $U = U_{\text{KK}}$, the period of the $\tau$ direction must be
\begin{align}
\delta\tau = \frac{4\pi}{3}\frac{R^{3/2}}{U_{\text{KK}}^{1/2}} = \frac{2\pi}{M_{\text{KK}}} \ .
\label{eq:tau}
\end{align}

\noindent
We define the Kaluza-Klein mass $M_{\text{KK}}$ by the right hand side of eq. (\ref{eq:tau}). 
$M_{\text{KK}}$ characterizes the energy scale below which the dual gauge theory can be effectively regarded as a four dimensional pure Yang-Mills theory, and physically it sets a dynamical scale which is treated as a ``QCD scale.''
The four dimensional Yang-Mills coupling $g_{\text{YM}}$ can be read off from the DBI action for the D4-brane compactified on the $S^{1}$ as $g_{\text{YM}}^{2} = (2\pi)^{2}g_{s}l_{s}/\delta\tau$. 
Then, the parameters $R$, $U_{\text{KK}}$, and $g_{s}$ are expressed in terms of gauge theory parameters: 
\begin{align}
R^{3} = \frac{1}{2}\frac{g_{\text{KK}}^{2}N_{c}l_{s}^{2}}{M_{\text{KK}}} \ , \quad U_{\text{KK}} = \frac{2}{9}g_{\text{YM}}^{2}N_{c}M_{\text{KK}}l_{s}^{2} \ , \quad g_{s} = \frac{1}{2\pi}\frac{g_{\text{YM}}^{2}}{M_{\text{KK}}l_{s}} \ . 
\end{align}

Let us consider here the embedding of a probe D6-brane in the D4 background. 
Introducing a new radial coordinate $\rho$ defined by $U(\rho) = \left( \rho^{3/2} + U_{\text{KK}}^{3}/4\rho^{3/2} \right)^{2/3}$, the D4 background is then
\begin{align}
ds^{2} = \left( \frac{U(\rho)}{R} \right)^{3/2}\left( \eta_{\mu\nu}dx^{\mu}dx^{\nu} + f(U)d\tau^{2} \right) + K(\rho)\left( d\lambda^{2} + \lambda^{2}d\Omega_{2}^{2} + dr^{2} + r^{2}d\phi^{2} \right) \label{eq:d4soliton2} \ , 
\end{align}

\noindent
where $K(\rho)=R^{3/2}U(\rho)^{1/2}/\rho^{2}$ and $\rho^{2} = \lambda^{2} + r^{2}$. 
Taking the static gauge, the induced metric on the D6-brane is, with ansatz given in \cite{Kruczenski}, 
\begin{align}
ds_{D6}^{2} = \left( \frac{U}{R} \right)^{3/2} \eta_{\mu\nu}dx^{\mu}dx^{\nu}+K\left[ (1+\dot{r}(\lambda)^{2})d\lambda^{2}+\lambda^{2}\Omega_{2}^{2} \right] \ , 
\end{align}

\noindent
where $\dot{r}=\partial_{\lambda}r$. 
The D6-brane action reads 
\begin{align}
S_{D6} &= -\frac{1}{(2\pi)^{6}l_{s}^{7}}\int d^{7}\sigma\, e^{-\phi}\sqrt{-\det g} \nonumber \\
 &= -T_{D6}\int d^{4}x d\Omega_{2}d\lambda \left( 1 + \frac{U_{\text{KK}}^{3}}{4\rho^{3}} \right)^{2}\, \lambda^{2} \sqrt{1+\dot{r}(\lambda)^{2}} \ ,
\label{eq:d6action} 
\end{align}
where $T_{D6}=2\pi/g_{s}(2\pi l_{s})^{7}$ is the D6-brane tension. 
The D6-brane configuration in the D4 background is determined by the equation of motion for $r(\lambda)$: 
\begin{align}
\frac{d}{d\lambda}\left[ \left( 1 + \frac{1}{4\rho^{3}} \right)^{2}\lambda^{2}\frac{\dot{r}}{\sqrt{1 + \dot{r}^{2}}} \right] 
= -\frac{3}{2}\frac{1}{\rho^{5}}\left( 1 + \frac{1}{4\rho^{3}} \right)\lambda^{2}\ r\sqrt{1 + \dot{r}^{2}} \ .
\label{eq:eomr}
\end{align}

\noindent
Here a redefinition $\lambda\to U_{\text{KK}}\lambda,\, r\to U_{\text{KK}}r,\, \rho\to U_{\text{KK}}\rho$,
introduces dimension-less coordinates.
The asymptotic behavior of the solution to eq. (\ref{eq:eomr}) is 
\begin{align}
r_{v}(\lambda) \simeq r_{\infty} + \frac{c}{\lambda} \, , \quad \text{for} \quad \lambda\to\infty \, , 
\end{align}

\noindent
where $r_{\infty}$ is the asymptotic distance between the D4-branes and the D6-brane, which is related to the quark mass $m_{q}$ as $m_{q} = U_{\text{KK}}r_{\infty}/2\pi l_{s}^{2}$.

To look at the low energy effective theory of mesons, we introduce fluctuations of the D6-brane around the static solution, which are the following embeddings of the D6-brane: 
\begin{align}
r = r_{v}(\lambda) + \delta r \, , \quad \phi = 0 + \delta\phi \, , \quad \tau = \text{constant} \, ,
\label{eq:randphi}
\end{align}

\noindent
where the fluctuations $\delta r$ and $\delta\phi$ are now functions of all the world-volume coordinates, and $r_{v}$ is the numerically-determined static solution to eq. (\ref{eq:eomr}). 
(Note that $r$ and $r_{v}$ in eq. (\ref{eq:randphi}) are the ones without the above rescaling.)
The induced metric on the D6-brane is then 
\begin{align}
ds^{2} &= \left( \frac{U}{R} \right)^{3/2}\eta_{\mu\nu}dx^{\mu}dx^{\nu} + K\left[ (1 + \dot{r}_{v}^{2})d\lambda^{2} + \lambda^{2}d\Omega_{2}^{2} \right] \nonumber \\
 & + 2K\dot{r}_{v}(\partial_{a}\delta r)d\lambda dx^{a} + K\left[ (\partial_{a}\delta r)(\partial_{b}\delta r) + (r_{v} + \delta r)^{2}(\partial_{a}\delta \phi)(\partial_{b}\delta \phi) \right] dx^{a}dx^{b} \, . 
\label{eq:d6flucmetric}
\end{align}

\noindent
The indices $a,\, b$ run over all the world-volume directions. 
The D6-brane action now becomes
\begin{align}
S &= -\frac{2\pi}{(2\pi l_{s})^{7}}\int d^{7}\sigma \, e^{-\phi}\sqrt{-\det \left( \tilde{g} + \delta g \right)} \nonumber \\
 &= -T_{D6} U_{\text{KK}}^{3}\int d^{4}xd\Omega_{2}d\lambda\, \lambda^{2}\sqrt{1+\dot{r}_{v}}\left( 1 + \frac{1}{4\rho^{3}} \right)^{2}\sqrt{\det \left( 1+\tilde{g}^{-1}\delta g \right)} \, . 
\label{eq:d6action2}
\end{align}

\noindent
$\tilde{g}$ denotes the first line of the metric (\ref{eq:d6flucmetric}) and $\delta g$ second line, and we have rescaled $\lambda,\, r,\, \rho$ as before. 
Expanding the D6-brane action (\ref{eq:d6action2}) to quadratic order in the fluctuations, we may find the equations of motion for $\delta r \, , \delta\phi$. 
By separating variables as 
\begin{align}
\delta r = R(\lambda) e^{ik_{r}\cdot x} Y_{l_{r}m_{r}}(\Omega_{2}) \, , \quad 
 \delta\phi = P(\lambda) e^{ik_{\phi}\cdot x}Y_{l_{\phi}m_{\phi}}(\Omega_{2}) \, , 
\end{align}

\noindent
where $Y_{lm}$ is a spherical harmonics on $S^{2}$, the equations of motion become eigenequations for $R$ and $P$. 
Focusing on the modes $l_{r,\phi} = 0$, the eigenequations for $R$ and $P$ yield the spectra $M_{r,\phi}^{2}$ and the eigenfunctions $R_{n},\, P_{n}$, where $n=0,1,2,\cdots$ labels the eigen states.
Since we are interested in the dynamics at low energy, we consider only the lowest mode,  $R_{0}$ and $P_{0}$, henceforth.

\subsection{Relation to the linear sigma model}
\label{sec:d4d6sigma}

The above D4/D6 system is a model of mesons in a 1 flavor holographic QCD, and it has the same structure of the chiral U$(1)_{A}$ symmetry and its breaking as the linear sigma model considered in section \ref{sec:linearsigma}. 
Thus, we regard the two systems as a same low energy effective theory of mesons.
To map the phase diagram of chaos obtained in section \ref{sec:linearsigma} to the D4/D6 model, 
in this subsection we compare the parameters of the models\footnote{For the details of the calculation, see appendix \ref{sec:calculation}. } and find explicit relations.

To determine the relation between the two systems, 
we have to consider not only mass terms, but also interaction terms. 
So, we choose a strategy to expand the D6-brane action (\ref{eq:d6action2}) and the linear sigma model action
to the cubic order in the fluctuations, and compare each term to relate the two.\footnote{This strategy does
not immediately certify  that higher order terms coincide with each other. In fact, in general they are different in the
two models, and in any low energy models. We implicitly assume that the quadratic and cubic terms capture the chaotic dynamics, because the D6-brane action suffers from infinite number of terms in fluctuations.}

Let us first work out the fluctuation expansion for the D6-brane action.
We separate the variables of the fluctuations as follows:
\begin{align}
\delta r = \mathcal{N}_{r} \, \delta r(t) R_{0}(\lambda) \ , \quad \delta\phi = \mathcal{N}_{\phi} \, \delta\phi(t) P_{0}(\lambda) \ . 
\label{eq:expansion}
\end{align}

\noindent
The normalizations $\mathcal{N}_{r,\phi}$ is determined to make kinetic terms canonical. 
Substituting (\ref{eq:expansion}) into the action and integrating over $x^{1,2,3},\, \lambda$, and $\Omega_{2}$, we find 
\begin{align}
S &= \int dt\bigg[ \frac{1}{2}\left( \left( \delta r'(t)\right)^{2}+\left( \delta \phi'(t)\right)^{2} \right) - M_{\text{KK}}^{2}\left( A(\delta r)^{2}+B(\delta \phi)^{2} \right) \nonumber \\
&\qquad\quad -\frac{1}{\lambda\sqrt{N_{c}}}\bigg\{ M_{\text{KK}}\left( C_{1} \delta r(\delta\phi )^{2}+C_{2}(\delta r)^{3}\right) \nonumber \\
&\qquad\qquad\qquad\quad +\frac{1}{M_{\text{KK}}}\left( D_{1}\delta r(\delta r')^{2}+D_{2}\delta r'\delta\phi'\delta\phi +D_{3}\delta r(\delta\phi')^{2}   \right)  \bigg\}
 \bigg] \ , 
\label{eq:d6flucaction}
\end{align}

\noindent
where the action is normalized by the spatial initegration $\int d^{3}x$, and $'$ is time derivative, $' = \partial_{t}$. 
Also, we have defined the 't Hooft coupling as $\lambda := g_{\text{YM}}^{2}N_{c}$. 
  $A,\, C,\, D$ and so on are just numerical coefficients.\footnote{Their explicit expressions are found in appendix
  \ref{sec:fluc}.}
The coefficients in front of $(\delta r)^{2}$ and $(\delta\phi)^{2}$ represent the mass for each field, and they are actually given by the spectra given in the previous subsection.

On the other hand,
the action for the linear sigma model considered in section \ref{sec:linearsigma} with spatially homogeneous fields is 
\begin{align}
S = \int dt \bigg[ \frac{1}{2}\left( \sigma'(t)^{2}+\pi'(t)^{2} \right) - \left( \frac{\mu^{2}}{2}\left( \sigma^{2}+\pi^{2} \right) +\frac{g_{4}}{4}\left( \sigma^{2}+\pi^{2} \right)^{2}+a\sigma \right)  \bigg] \ .
\label{eq:linearsgaction}
\end{align}

\noindent
This is also normalized by $\int d^{3}x$, and $' = \partial_{t}$. 
Let us introduce a polar coordinate, $\sigma = R\cos\theta,\, \pi = R\sin\theta$, 
and field redefinitions around the static solution as
\begin{align}
R &= f_{\pi} + \delta R + \epsilon_{1}\left( \delta R \right)^{2} + \epsilon_{2}\left( \delta\theta \right)^{2} + \epsilon_{3}\left( \delta R' \right)^{2} \ , \label{eq:redefinition1} \\
\theta &= 0 + \delta\theta + \epsilon_{4}\delta R \delta\theta \ . \label{eq:redefinition2}
\end{align}

\noindent
This redefinition is chosen in the following way. 
In order to respect the invariance of the action under the parity transformation $\delta\theta \to -\delta\theta$, we consider redefinitions by which an odd order of $\delta\theta$ does not appear in the action. 
In addition, we choose redefinitions to make the same form of the action as the one of the D6-brane (\ref{eq:d6flucaction}) up to cubic order. 
The terms with derivative are included only if the Lagrangian can be reduced to the same form as eq. (\ref{eq:d6flucaction}) by partial integration. 
When the quark is massless, the rotation (chiral) symmetry of the linear sigma model is restored; that is, when $a\to 0$, $\epsilon_{2}$ and $\epsilon_{4}$ must be $0$ due to the symmetry. 
This can be confirmed by numerical analysis after the comparison with the D4/D6 system.\footnote{This field redefinitions still have ambiguity. The details are discussed in appendix \ref{sec:redefinition}.}

By this redefinitions, renormalizing $f_{\pi}\delta\theta \to \delta\theta$, up to cubic order in the fluctuations we find
\begin{align}
S &= \int dt\bigg[ \frac{1}{2}\bigg( (\delta R'(t))^{2} + (\delta\theta'(t))^{2}\bigg)-\frac{1}{2}\left( \mu^{2}+3g_{4}f_{\pi}^{2} \right)(\delta R)^{2} + \frac{a}{2f_{\pi}}(\delta\theta)^{2} \nonumber \\
 &\quad + \left( \frac{a}{2f_{\pi}^{2}} - \frac{\mu^{2}\epsilon_{2}}{f_{\pi}^{2}} -3g_{4}\epsilon_{2} + \frac{a\epsilon_{4}}{f_{\pi}} \right) \delta R(\delta\theta)^{2} - \left( g_{4}f_{\pi} + 3g_{4}f_{\pi}^{2}\epsilon_{1}+\mu^{2}\epsilon_{1} \right)(\delta R)^{3} \nonumber \\
 &\quad + \left( 2\epsilon_{1} - \mu^{2}\epsilon_{3}-3g_{4}f_{\pi}^{2}\epsilon_{3} \right) \delta R(\delta R')^{2} + \left( \frac{2\epsilon_{2}}{f_{\pi}^{2}} + \epsilon_{4} \right) \delta R'\delta\theta'\delta\theta + \left( \frac{1}{f_{\pi}} + \epsilon_{4} \right)\delta R(\delta\theta')^{2}  \bigg] \ . 
\label{eq:linearsgflucaction}
\end{align}

\noindent
Identifying the coefficients in front of each term in the actions (\ref{eq:d6flucaction}) and (\ref{eq:linearsgflucaction}) yields seven equations. 
We can uniquely solve the equations for $\mu^{2},\, g_{4},\, a,\, \epsilon_{1,2,3,4}$ with the equation of motion for the static solution in the linear sigma model (\ref{eq:fpi}), 
which determines $f_{\pi}$. 
Then, the parameters in the linear sigma model $\mu^{2},\, g_{4},\, a$ are expressed in terms of gauge theory parameters in the D4/D6 system, $\lambda,\, N_{c},\, M_{\text{KK}}$ as 
\begin{align}
\mu^{2} = -0.39\, M_{\text{KK}}^{2}\, , \quad g_{4} = 7.9\times 10^{3}\, \frac{1}{\lambda^{2}N_{c}}\, , \quad a = - 7.0\times 10^{-5}\, M_{\text{KK}}^{3}\lambda\sqrt{N_{c}} \ .
\label{eq:matching}
\end{align}
This is the relation between the linear sigma model and the gauge theory parameters, via the holographic correspondence.

The dependence on $r_{\infty}$ is included in the numerical coefficients, since $r_{\infty}$ is a boundary condition imposed on the static configuration of the D6-brane, $r_{v}$. 
The numerical coefficients shown in eqs. (\ref{eq:matching}) are calculated at $r_{\infty} = 0.015$. 
Actually, $r_{\infty}$ corresponds to the quark mass, and we can find by numerical analysis that the numerical coefficient of $a$ is linearly dependent on $r_{\infty}$, which is consistent. 
In particular, when $r_{\infty} = 0$, the coefficient of $a$ is also $0$, for sure, and then the rotation symmetry of the linear sigma model is restored.

\subsection{A phase diagram of chaos: $\lambda,\, N_{c}$ dependence of QCD chaos}
\label{sec:chaosviad4d6}

Using the relations (\ref{eq:matching}), we can explore the chaos of the linear sigma model with changing the parameters $\lambda,\, N_{c},\, M_{\text{KK}}$, and $r_{\infty}$. 
As in the case of the linear sigma model, 
we here show that the system has a specific scaling law for $\lambda$ and $N_{c}$. 
It will reduce the number of relevant parameters and helps us drawing the phase diagram of chaos.

Let us rewrite the relations (\ref{eq:matching}) as 
\begin{align}
\mu^{2} = \mu^{2}(r_{\infty})\, M_{\text{KK}}^{2}\, , \quad g_{4} = g_{4}(r_{\infty})\, \frac{1}{\lambda^{2}N_{c}}\, , \quad a = a(r_{\infty})\, M_{\text{KK}}^{3}\lambda\sqrt{N_{c}} \ .
\end{align}

\noindent
Then, the Hamiltonian of the linear sigma model is written as
\begin{align}
H = \frac{1}{2}(p_{\sigma}^2 + p_{\pi}^2) + \frac{\mu^{2}(r_{\infty}) M_{\text{KK}}^2}{2}(\sigma^2 + \pi^2) + \frac{1}{4}\frac{g_4(r_{\infty})}{\lambda^2 N_c}(\sigma^2 + \pi^2)^2 + a(r_{\infty})M_{\text{KK}}^3\lambda\sqrt{N_c}\, \sigma \ .
\end{align}

\noindent
Under the rescaling 
\begin{align}
\sigma\to \lambda\sqrt{N_{c}}M_{\text{KK}}\sigma \, , \quad \pi\to \lambda\sqrt{N_{c}}M_{\text{KK}}\pi \, , \quad t\to M_{\text{KK}}^{-1}t \ ,
\end{align}

\noindent
the Hamiltonian becomes
\begin{align}
\frac{H}{\lambda^{2}N_{c}M_{\text{KK}}^{4}} = \frac{1}{2}(p_{\sigma}^2 + p_{\pi}^2) + \frac{\mu^{2}(r_{\infty})}{2}(\sigma^2 + \pi^2) + \frac{g_4(r_{\infty})}{4}(\sigma^2 + \pi^2)^2 + a(r_{\infty})\, \sigma \ .
\label{eq:scaleinvhamiltonian}
\end{align}

\begin{figure}[tbp]
\centering
\includegraphics[width=100mm,bb=0 0 1024 768,trim=120 100 0 100]{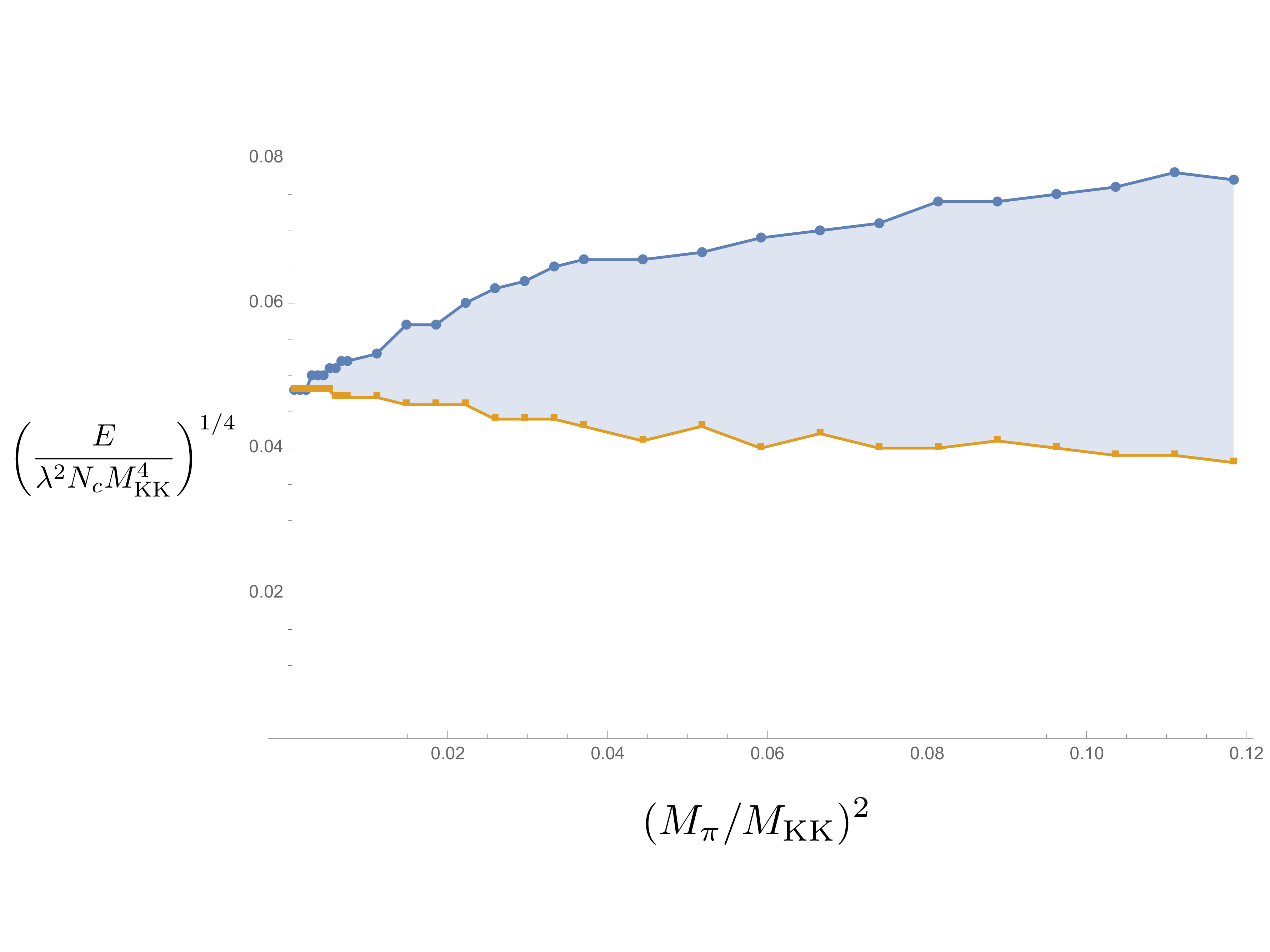}
\caption{The phase diagram of chaos in the linear sigma model defined by the Hamiltonian (\ref{eq:scaleinvhamiltonian}). The horizontal axis is $(M_{\pi}/M_{\text{KK}})^{2} = 0.74\, r_{\infty}$. In the shaded region, the system exhibis chaos. At each energy, initial conditions are chosen as $(\pi,\, p_{\sigma},\, p_{\pi}) = (0.05 f_{\pi} \times i,\, 0,\, 0)$, $i = 1,\dots,7$, and $(\sigma,\, p_{\sigma},\, p_{\pi}) = (0.05 f_{\pi} \times j,\, 0,\, 0)$, $j = \pm 1,\dots,\pm 10$. }
\label{fig:diagramrinfty}
\end{figure}

\noindent
Therefore, changing $\lambda,\, N_{c}$ means the rescaling of the coordinates $\sigma$ and $\pi$. 
In particular, $\lambda$ and $N_{c}$ appear only in a combination $\lambda\sqrt{N_{c}}$, which implies that the system has a specific scaling law under the rescaling of  $\lambda\sqrt{N_{c}}$. 
Equation (\ref{eq:scaleinvhamiltonian}) shows that only $r_{\infty}$ changes the model. 
In contrast to the analysis in section \ref{sec:linearsigma}, this time we have one parameter $r_{\infty}$, which characterizes the system.\footnote{ 
A difference from the previous section is that $r_{\infty}$ changes not only $a(r_{\infty})$, but also $\mu^{2}(r_{\infty})$ and $g_{4}(r_{\infty})$.}
Thus, we conclude that the dynamics of the single flavor large 
$N_c$ QCD at low energy with different $\lambda,\, N_{c}$ are related each other by the specific scaling transformation, and the quark mass, or the pion mass characterizes the theory.

Based on this scaling argument, we investigate the chaos of the linear sigma model with the parameter $r_{\infty}$. 
Similarly to section \ref{sec:linearsigma}, by numerically calculating the Poincar\'e sections in detail, we make a judge whether chaos appears or not at each energy and at each value of $r_\infty$. 
Using the relation between $r_{\infty}$ and the pion mass, $M_{\pi}^{2} = 0.74M_{\text{KK}}^{2}r_{\infty}$,\footnote{This relation is given in \cite{Kruczenski}, section 3.1. } the phase diagram of chaos in the linear sigma model is now rephrased in terms of $r_\infty$, $\lambda$ and $N_c$. See figure \ref{fig:diagramrinfty}, the phase
diagram of chaos for the D4/D6 holographic QCD model.
The horizontal axis and the vertical axis are chosen as the scale-invariant combinations. 
In the shaded region the system is chaotic, and in others chaos does not appear.

Figure \ref{fig:diagramrinfty} shows that the system is more chaotic for larger $r_{\infty}$, i.e. larger quark mass, or larger pion mass. 
This is qualitatively same as the result in section \ref{sec:linearsigma}. 
However, the significant difference from the previous section is that the dependence of the chaos on $\lambda$ and $N_{c}$ is uncovered. 
The horizontal and vertical axis in figure \ref{fig:diagramrinfty} are chosen as scale-invariant combinations, so we can read off from figure \ref{fig:diagramrinfty} that we need more energy to cause chaos in this system, as larger $\lambda$ or larger $N_{c}$. 
In other words, it is hard to cause chaos in a strongly coupled gauge theory with larger $N_{c}$. 

This is a totally counterintuitive result, since larger $\lambda$ means the gauge theory is strongly coupled and larger $N_{c}$ means the degrees of freedom is larger. 
From the classical or perturbative point of view, a gauge theory should be more chaotic with stronger coupling and with more degrees of freedom, but our result implies that a non-perturbative quantum gauge theory may be less chaotic with stronger coupling and with more gluonic degrees of freedom.

\begin{figure}[tbp]
\centering
\includegraphics[width=100mm,bb=0 0 1024 768,trim=120 100 0 100]{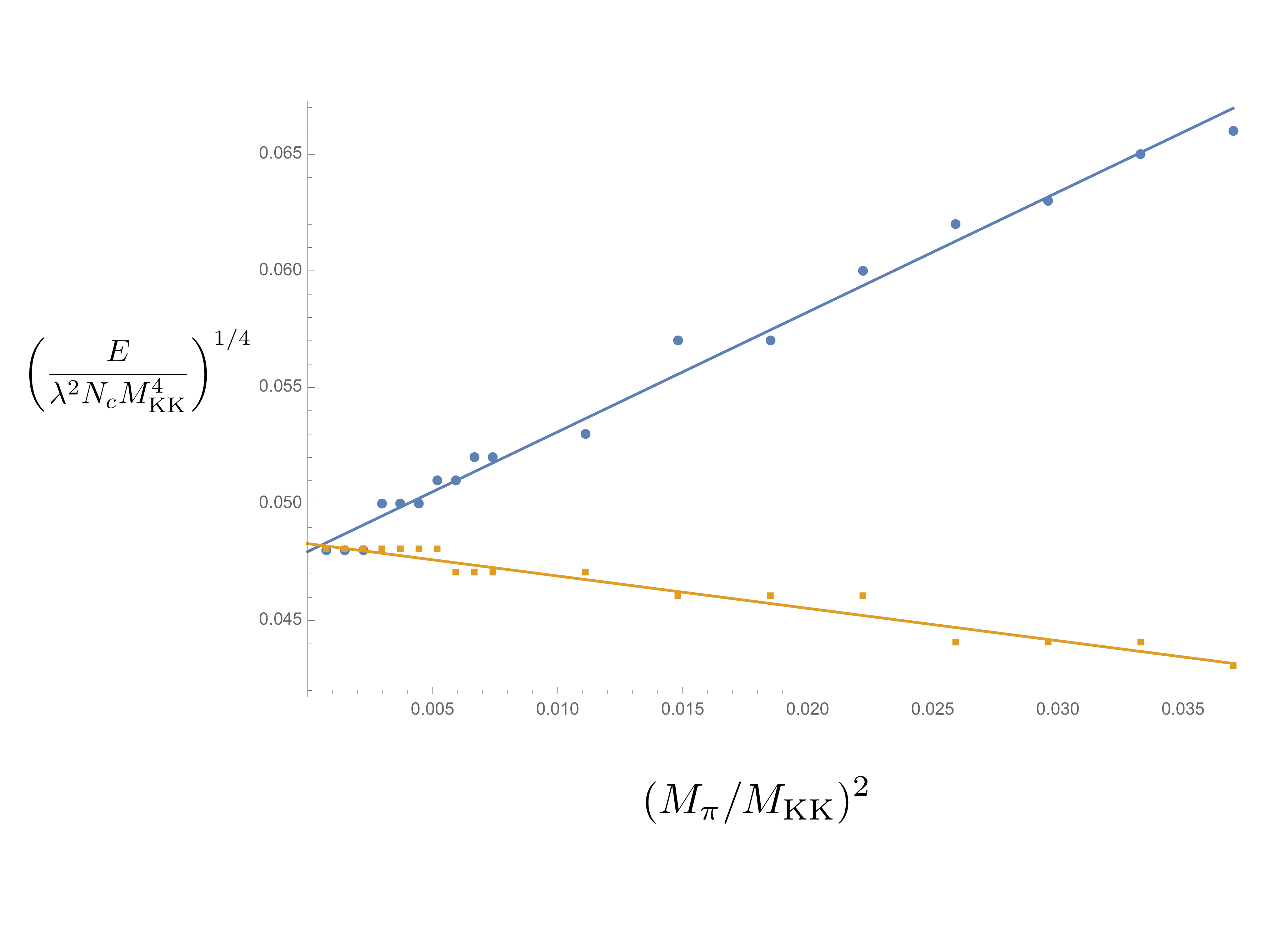}
\caption{The linear fitting in figure \ref{fig:diagramrinfty} from $r_{\infty} = 0$ to $0.05$. The horizontal axis is again $(M_{\pi}/M_{\text{KK}})^{2} = 0.74\, r_{\infty}$. Our numerical analysis shows that the slope of the upper bound is approximately $A = 0.514$, and the one of the lower bound is $-\tilde{A} = -0.139$. Both of the intercept at $r_{\infty} = 0$ can be regarded as an identical value within the numerical error. }
\label{fig:linearfitrinfty}
\end{figure}

Finally, let us study the chaos phase boundary and its dependence on $\lambda$ and $N_{c}$ in some detail. 
Figure \ref{fig:linearfitrinfty} shows a linear fitting of the phase boundaries in figure \ref{fig:diagramrinfty} using
the data point in the range  $0\leq r_{\infty} \leq 0.05$. 
In this range, the upper bound of the chaos and the lower bound can be expressed as the following linear equations: 
\begin{align}
\left( \frac{E_{u}}{\lambda^{2}N_{c}M_{\text{kk}}^{4}} \right)^{1/4} &= A \left( 0.74\, r_{\infty} \right) + b \ , \quad A>0 \ , \\
\left( \frac{E_{l}}{\lambda^{2}N_{c}M_{\text{kk}}^{4}} \right)^{1/4} &= -\tilde{A} \left( 0.74\, r_{\infty} \right) + b\ , \quad \tilde{A}>0 \ .
\end{align}
Using the relation $r_{\infty} = 9\pi m_{q}/\lambda M_{\text{KK}}$,\footnote{This relation is given in \cite{Kruczenski}, section 3.1. } the above equations imply that the energy region of the occurrence of the
chaos depends on $\lambda$ and $N_{c}$ at each quark mass as 
\begin{align}
E_{u}^{1/4} - E_{l}^{1/4} = 13.7 \times m_{q} \frac{N_{c}^{1/4}}{\sqrt{\lambda}} \ .
\label{eq:energywidth}
\end{align}

\noindent
This means that the width of the energy in which the system exhibits chaos is wider for larger $N_{c}$, and is narrower for larger $\lambda$. 
Note that, in contrast, the magnitude of the energy itself at which the system exhibits chaos increases as larger $\lambda$ and larger $N_{c}$.
This difference stems from the relation between the quark mass and the pion mass in which $\lambda$ shows up.



\section{Conclusion and discussion}
\label{sec:conclusion}

In this paper, 
we have evaluated the chaos of the linear sigma model and the D4/D6 system, and have drawn a phase diagram of
chaos. Figure \ref{sigmaphasediagram1} and figure \ref{fig:diagramrinfty} are our results of the phase diagram.
They have a simple structure: at a vanishing quark mass the chaos disappears, while with an infinitesimally small
quark mass the chaos appears, and grows.
The linear sigma model is more chaotic for larger quark mass, or pion mass.
Using the D4/D6 model, we have recast the parameters in the linear sigma model to those of the gauge theory,
and have found that
in a non-perturbative quantum gauge theory it is harder for the chaos to occur for larger $\lambda$ and larger $N_{c}$. 

The latter is counterintuitive, but technically easy to understand, since
from the view point of the AdS/CFT, in the gravity dual $\lambda$ and 
$N_{c}$ appear as in $1/\lambda$ or $1/N_{c}$ expansion. 
Therefore, our result would never be seen in any perturbative analysis of QCD.
Interestingly, as for $\lambda$ and $N_c$, we found that the low energy effective theories of the single 
flavor QCD with different $\lambda,\, N_{c}$ are related each other by a specific scaling transformation, and 
only the quark mass, or the pion mass characterizes the theory.

In section \ref{sec:linearsigma}, we have investigated the chaos of the linear sigma model in detail, and found that the quark mass term which causes the explicit breaking of the chiral symmetry is crucial to chaos.
However, figure \ref{sigmaphasediagram3} shows that the chaos appears not from the saddle point associated with the quark mass term, but from the top of the potential, i.e. the maximal point, $(\sigma,\, \pi) \simeq (0,\, 0)$. 
It is interesting to interpret this fact as a relation to the standard QCD phase diagram. As we have discussed in
section \ref{sec:linearsigma}, the QCD thermal phase transition, together with an entropy production, 
takes place at QCD scale, irrespective to the quark mass. On the other hand, the saddle point energy vanishes
when the quark mass is zero, which does not share the same property as the thermal phase transition energy
scale. The maximum point is the one to share the same property, and we found that the chaos originates there.
It would be nice if we can trace more on possible relations between the thermal phase transition of QCD and
our chaos phase diagram. Numerical evaluation of Lyapunov exponents may provide a key from the viewpoint of entropies.

To draw the phase diagram of chaos, we relied on the accidental scaling law we discovered which appears
in the combination $\lambda\sqrt{N_{c}}$. 
The analyses would have been more complicated if it were not for
the scaling law.
Actually, if we instead consider the multi-flavor case, $N_{f}$ D6-branes ($N_{f}>1$), or vector mesons, 
the combination $\lambda\sqrt{N_{c}}$ would not be universal. 
It would be interesting to further look at the reason why the combination appears in the effective actions,
and how generic our scaling law can be found in holographic models.
Popular models such as D4/D8 system \cite{Sakai1,Sakai2} may be a good starting point for the study.

A relating comment on the $\lambda$ and $N_c$ dependence is about our eq. (\ref{eq:energywidth}).
It shows that the energy range in which the system exhibits chaos is proportional to $m_{q}N_{c}^{1/4}/\sqrt{\lambda}$. 
This coincides exactly with the chaos energy range obtained in \cite{Hashimoto}, 
for the $N_f=2$ supersymmetric holographic QCD model.
It may be interesting to study this coincidence in more detail, using other codimension four brane set up, for example the D1/D5 system, the D5/D9 system, and so on. 

A technical assumption we made in the mapping between the linear sigma model and the 
D4/D6 system is that we compare only low order terms. 
In general, although low energy models are different in higher order terms, 
they are believed to
provide similar physics at low energy. That is why we have compared 
only quadratic and cubic terms of the two models
to relate them. However, infinitely many 
higher order terms in the D4/D6 system may affect the chaos analysis. For example,
the D6-brane is not allowed to enter the spacetime region $U \leq U_{\text{KK}}$ due to the topology
of the D4 background, so the saddle point is expected to be present for any value of the quark mass,
while the linear sigma model does not have that property, see figure \ref{sigmaphasediagram3}. 
The full D6-brane action suffers from infinite number of higher order fluctuation terms, so
it is a technical challenge to solve the full motion of the D6-brane. It is natural to believe that
the topological structure of the phase diagram of chaos will not change drastically by the inclusion of
the higher order terms, and we here leave it as a future problem.

We have obtained ``a phase diagram of QCD chaos,'' which is a map of chaos in QCD-like gauge theories 
at low energy. 
We have briefly mentioned its possible relation to the standard QCD phase diagram, but the 
explicit relations between our phase diagram of chaos and the standard QCD phase diagram are yet 
to be explored. Since chaos involves time-dependent evolution of the states,
non-equilibrium steady states in QCD rather than static states used for drawing the standard QCD
phase diagram would be necessary to relate the two. A further study, together with the AdS/CFT correspondence,
will lead us to a more unified picture of the phase diagrams.

\acknowledgments

We would like to thank Keiju Murata for valuable discussions.
The work of K.H. was supported in part by JSPS KAKENHI
Grants No.~JP15H03658, No.~JP15K13483, and
No.~JP17H06462.

\clearpage

\appendix
\section{Calculations in the comparison between the linear sigma and the D4/D6}
\label{sec:calculation}

We here give detailed calculations used in section \ref{sec:D4D6}. 
We first write down the D6-brane DBI action expanded to cubic order in the fluctuations, and second we show the comparison between the linear sigma model and the D4/D6 system. 
Then, the parameters in the linear sigma model are expressed by the ones in the D4/D6 system, and we can confirm the restoration of the rotation symmetry by numerical analysis.

\subsection{Expansion of the D6-brane action to cubic order in fluctuations}
\label{sec:fluc}

Let us expand the D6-brane action (\ref{eq:d6flucaction}) to cubic order in the fluctuations, and substitute $\delta r = \mathcal{N}_{r}\, \delta r(t) R_{0}(\lambda),\, \delta\phi = \mathcal{N}_{\phi}\, \delta\phi(t) P_{0}(\lambda)$ into it. Then, integrating over $x^{1,2,3},\, \lambda$, and $\Omega_{2}$, the action turns out to be
\begin{align}
S &= 4\pi T_{D6}U_{\text{KK}}^{3}\int dt \bigg[\frac{R^{3}}{U_{\text{KK}}}\bigg(\frac{\mathcal{N}_{r}^{2}}{2}  \int d\lambda \left( 1+\frac{1}{4\rho_{v}^{3}} \right)^{2}\frac{\lambda^{2}R_{0}(\lambda)^{2}}{\rho_{v}^{2}U_{v}\sqrt{1+\dot{r}_{v}^{2}}} \left( \delta r'(t) \right)^{2} \nonumber \\
&\quad\quad\quad\quad\quad\quad\quad\quad\quad\quad\quad +\frac{\mathcal{N}_{\phi}^{2}}{2} \int d\lambda \left( 1+\frac{1}{4\rho_{v}^{3}} \right)^{2}\frac{\lambda^{2}r_{v}^{2}\sqrt{1+\dot{r}_{v}^{2}}P_{0}(\lambda)^{2}}{\rho_{v}^{2}U_{v}} \left( \delta\phi'(t) \right)^{2}  \bigg) \nonumber \\
& - \mathcal{N}_{r}^{2}\int d\lambda\, \lambda^{2}\sqrt{1+\dot{r}_{v}^{2}}\bigg\{  \left( 1+\frac{1}{4\rho_{v}^{3}} \right)^{2}\frac{\dot{R}_{0}^{2}}{2(1+\dot{r}_{v}^{2})^{2}}+\left( \frac{3(7r_{v}^{2}-\lambda^{2})}{16\rho_{v}^{10}}+\frac{3(4r_{v}^{2}-\lambda^{2})}{4\rho_{v}^{7}} \right) R_{0}^{2} \nonumber \\
&\quad\quad\quad\quad\quad\quad\quad\quad\quad\quad\quad\quad -\left( 1+\frac{1}{4\rho_{v}^{3}} \right)\frac{3r_{v}\dot{r}_{v}}{2\rho_{v}^{5}(1+\dot{r}_{v}^{2})}R_{0}\dot{R}_{0} 
   \bigg\} (\delta r)^{2} \nonumber \\
& -\mathcal{N}_{\phi}^{2}\int d\lambda \left( 1+\frac{1}{4\rho_{v}^{3}} \right)^{2}\frac{\lambda^{2}r_{v}^{2}\dot{P}_{0}^{2}}{2\sqrt{1+\dot{r}_{v}^{2}}} (\delta\phi)^{2} \nonumber \\
& -\mathcal{N}_{r}\mathcal{N}_{\phi}^{2}\int d\lambda\, \lambda^{2}\sqrt{1+\dot{r}_{v}^{2}}\bigg\{ \left( \left( 1+\frac{1}{4\rho_{v}^{3}} \right)^{2} +\frac{3r_{v}^{2}}{4\rho_{v}^{5}}\left( 1+\frac{1}{4\rho_{v}^{3}} \right) \right) \frac{r_{v}R_{0}\dot{P}_{0}}{1+\dot{r}_{v}^{2}} \nonumber \\
&\quad\quad\quad\quad\quad\quad\quad\quad\quad\quad\quad\quad -\left( 1+\frac{1}{4\rho_{v}^{3}} \right)^{2} \frac{r_{v}^{2}\dot{r}_{v}\dot{R}_{0}\dot{P}_{0}^{2}}{2(1+\dot{r}_{v}^{2})^{2}}  \bigg\}  \left(\delta r(\delta\phi)^{2}  \right) \nonumber \\
& -\mathcal{N}_{r}^{3}\int d\lambda\, \lambda^{2}\sqrt{1+\dot{r}_{v}^{2}}\bigg\{ -\left( 1+\frac{1}{4\rho_{v}^{3}} \right)^{2}\frac{\dot{r}_{v}\dot{R}_{0}^{3}}{2(1+\dot{r}_{v}^{2})^{3}}+\frac{3r_{v}}{2\rho_{v}^{5}}\left( 1+\frac{1}{4\rho_{v}^{3}} \right) \frac{R_{0}\dot{R}_{0}^{2}}{2(1+\dot{r}_{v}^{2})} \nonumber \\
&\quad +\left( \frac{3(7r_{v}^{2}-\lambda^{2})}{16\rho_{v}^{10}}+\frac{3(4r_{v}^{2}-\lambda^{2})}{4\rho_{v}^{7}} \right) \frac{\dot{r}_{v}R_{0}^{2}\dot{R}_{0}}{1+\dot{r}_{v}^{2}}-\left( \frac{r_{v}(7r_{v}^{2}-3\lambda^{2})}{2\rho_{v}^{12}}+\frac{5r_{v}(4r_{v}^{2}-3\lambda^{2})}{4\rho_{v}^{9}} \right) R_{0}^{3}
   \bigg\}  (\delta r)^{3} \nonumber \\
&-\frac{R^{3}}{U_{\text{KK}}}\bigg( \mathcal{N}_{r}^{3}\int d\lambda\, \frac{\lambda^{2}}{2\rho_{v}^{3}\sqrt{1+\dot{r}_{v}^{2}}}\left( 1+\frac{1}{4\rho_{v}^{3}} \right)^{4/3}\bigg\{ \frac{r_{v}(7+12\rho_{v}^{3})R_{0}^{3}}{\rho_{v}^{2}(1+4\rho_{v}^{3})}+\frac{\dot{r}_{v}\dot{R}_{0}R_{0}^{2}}{1+\dot{r}_{v}^{2}}  \bigg\} \left( \delta r(\delta r')^{2} \right) \nonumber \\
&\quad +\mathcal{N}_{r}\mathcal{N}_{\phi}^{2}\int d\lambda \left( 1+\frac{1}{4\rho_{v}^{3}} \right)^{4/3}\frac{\lambda^{2}r_{v}^{2}\dot{r}_{v}R_{0}P_{0}\dot{P}_{0}}{\rho_{v}^{3}\sqrt{1+\dot{r}_{v}^{2}}} \left( \delta r'\delta\phi' \delta\phi \right) \nonumber \\
&\quad +\mathcal{N}_{r}\mathcal{N}_{\phi}^{2}\int d\lambda \frac{-\lambda^{2}r_{v}\sqrt{1+\dot{r}_{v}^{2}}}{2\rho_{v}^{2}U_{v}}\left( 1+\frac{1}{4\rho_{v}^{3}} \right)^{2}\bigg\{ \left( 2 - \frac{3r_{v}^{2}}{\rho_{v}^{2}}+\frac{r_{v}^{2}}{2\rho_{v}^{7/2}U_{v}^{3/2}} -\frac{6r_{v}^{2}}{\rho_{v}^{2}(1+4\rho_{v}^{3})}   \right) R_{0}P_{0}^{2} \nonumber \\
&\quad\quad\quad\quad\quad\quad\quad\quad\quad\quad\quad\quad +\frac{r_{v}\dot{r}_{v}\dot{R}_{0}P_{0}^{2}}{1+\dot{r}_{v}^{2}} \bigg\} \left( \delta r(\delta\phi')^{2} \right)
   \bigg)
    \bigg] 
    \nonumber
\end{align}

\begin{align}
 &= 4\pi T_{D6}U_{\text{KK}}^{3}\int dt\bigg[ \frac{R^{3}}{U_{\text{KK}}}\left( \frac{\mathcal{N}_{r}^{2}I_{r}}{2}\left( \delta r'(t)\right)^{2}+\frac{\mathcal{N}_{\phi}^{2}I_{\phi}}{2}\left( \delta \phi'(t)\right)^{2} \right) \nonumber \\
&\quad -\bigg\{ \mathcal{N}_{r}^{2}I_{A}(\delta r)^{2}+\mathcal{N}_{\phi}^{2}I_{B}(\delta\phi)^{2} +\mathcal{N}_{r}\mathcal{N}_{\phi}^{2}I_{C_{1}}\left( \delta r(\delta\phi )^{2}\right) +\mathcal{N}_{r}^{3}I_{C_{2}}(\delta r)^{3} \nonumber \\
&\quad +\frac{R^{3}}{U_{\text{KK}}}\left( \mathcal{N}_{r}^{3}I_{D_{1}}\left( \delta r(\delta r')^{2} \right) +\mathcal{N}_{r}\mathcal{N}_{\phi}^{2}I_{D_{2}}(\delta r'\delta\phi'\delta\phi) + \mathcal{N}_{r}\mathcal{N}_{\phi}^{2}I_{D_{3}}\left( \delta r(\delta\phi')^{2} \right) \right)   \bigg\}
  \bigg] \ , \label{eq:d6flucaction2}
\end{align}

\noindent
where again $\dot{} = \partial_{\lambda}$ and $' = \partial_{t}$, and the action is normalized by the spatial integration $\int d^{3}x$. 
$\rho_{v} := \lambda^{2} + r_{v}^{2}$ and $U_{v} = U(\rho_{v})$. 
$I$s are numerical coefficients given by the each integration over $\lambda$. 
Defining the 't Hooft coupling as $\lambda = g_{\text{YM}}^{2}N_{c}$, the normalizations $\mathcal{N}_{r,\phi}$ are determined to make the kinetic terms canonical: 
\begin{align}
\mathcal{N}_{r,\phi} = \left( 4\pi T_{D6} R^{3} U_{\text{KK}}^{2} I_{r,\phi} \right)^{-1/2} 
 = \frac{18\pi^{2}}{\sqrt{I_{r,\phi}}}\frac{1}{M_{\text{KK}}\lambda\sqrt{N_{c}}} \ . 
\end{align}

\noindent
With this normalizations, the D6-brane action finally becomes 
\begin{align}
S &= \int dt\bigg[ \frac{1}{2}\left( \left( \delta r'(t)\right)^{2}+\left( \delta \phi'(t)\right)^{2} \right) - M_{\text{KK}}^{2}\left( A(\delta r)^{2}+B(\delta \phi)^{2} \right) \nonumber \\
&\qquad\quad -\frac{1}{\lambda\sqrt{N_{c}}}\bigg\{ M_{\text{KK}}\left( C_{1} \delta r(\delta\phi )^{2}+C_{2}(\delta r)^{3}\right) \nonumber \\
&\qquad\qquad\qquad\quad +\frac{1}{M_{\text{KK}}}\left( D_{1}\delta r(\delta r')^{2}+D_{2}\delta r'\delta\phi'\delta\phi +D_{3}\delta r(\delta\phi')^{2}   \right)  \bigg\}
 \bigg] \ , 
 \end{align}

\noindent
where the coefficients are given by
\begin{align}
A &= \frac{4I_{A}}{9I_{r}} \, , \quad B = \frac{4I_{B}}{9I_{\phi}} \\
C_{1} &= \frac{8\pi^{2}I_{C_{1}}}{\sqrt{I_{r}}I_{\phi}} \, , \quad C_{2}=\frac{8\pi^{2}I_{C_{2}}}{\sqrt{I_{r}}I_{r}} \\
D_{1} &= \frac{18\pi^{2}I_{D_{1}}}{\sqrt{I_{r}}I_{r}} \, , \quad D_{2}=\frac{18\pi^{2}I_{D_{2}}}{\sqrt{I_{r}}I_{\phi}} \, , \quad D_{3}=\frac{18\pi^{2}I_{D_{3}}}{\sqrt{I_{r}}I_{\phi}} \ .
\end{align}

\subsection{Matching the parameters}

Identifying the coefficients in front of each term in the actions (\ref{eq:d6flucaction}) and (\ref{eq:linearsgflucaction}), we obtain the following seven equations: 
\begin{align}
M_{\text{KK}}^{2}A &= \frac{1}{2}\left( \mu^{2}+3g_{4}f_{\pi}^{2} \right) \ , \\
M_{\text{KK}}^{2}B &= -\frac{a}{2f_{\pi}} \ , \\
\frac{M_{\text{KK}}}{\lambda\sqrt{N_{c}}}C_{1} &= -\left( \frac{a}{2f_{\pi}^{2}} -\frac{\mu^{2}\epsilon_{2}}{f_{\pi}^{2}} -3g_{4}\epsilon_{2}+\frac{a\epsilon_{4}}{f_{\pi}} \right) \ , \\
\frac{M_{\text{KK}}}{\lambda\sqrt{N_{c}}}C_{2} &= g_{4}f_{\pi}+3g_{4}f_{\pi}^{2}\epsilon_{1}+\mu^{2}\epsilon_{1} \ , \\
\frac{1}{M_{\text{KK}}\lambda\sqrt{N_{c}}}D_{1} &= -\left( 2\epsilon_{1} -\mu^{2}\epsilon_{3}-3g_{4}f_{\pi}^{2}\epsilon_{3} \right) \ , \\
\frac{1}{M_{\text{KK}}\lambda\sqrt{N_{c}}}D_{2} &= -\left( \frac{2\epsilon_{2}}{f_{\pi}^{2}}+\epsilon_{4} \right) \ , \\
\frac{1}{M_{\text{KK}}\lambda\sqrt{N_{c}}}D_{3} &= -\left( \frac{1}{f_{\pi}}+\epsilon_{4} \right) \ .
\end{align}

\noindent
Besides them, we have the equation of motion for the static solution in the linear sigma model, which determines $f_{\pi}$, 
\begin{align}
\mu^{2}f_{\pi}+g_{4}f_{\pi}^{3}+a=0 \ .
\end{align}

\noindent
These eight equations uniquely determine $f_{\pi},\, \mu^{2},\, g_{4},\, a,\, \epsilon_{1,2,3,4}$ as
\begin{align}
f_{\pi} &= 7.1\times 10^{-3} \, M_{\text{KK}}\lambda\sqrt{N_{c}} \ , \\
\mu^{2} &= -0.39 \, M_{\text{KK}}^{2} \ , \\
g_{4} &= 7.9\times 10^{3} \, \frac{1}{\lambda^{2}N_{c}} \ , \\
a &= 7.0\times 10^{-5} \, M_{\text{KK}}^{3}\lambda\sqrt{N_{c}} \ , \\[2pt]
\epsilon_{1} &= -70 \, \frac{1}{M_{\text{KK}}\lambda\sqrt{N_{c}}} \ , \\
\epsilon_{2} &= 1.4\times 10^{-4} \, M_{\text{KK}}\lambda\sqrt{N_{c}} \ , \\
\epsilon_{3} &= 173 \, \frac{1}{M_{\text{KK}}^{3}\lambda\sqrt{N_{c}}} \ , \\
\epsilon_{4} &= -6.1 \, \frac{1}{M_{\text{KK}}\lambda\sqrt{N_{c}}} \ .
\end{align}
The numerical coefficients here are again calculated at $r_{\infty} = 0.015$. 
By numerical analysis, we find that the numerical coefficients of $\epsilon_{2}$ and $\epsilon_{4}$ tend to $0$ at $r_{\infty} \to 0$ as well as the one of $a$, which is in agreement with the restoration of the rotation symmetry.

\section{Field redefinition in the linear sigma model}
\label{sec:redefinition}

In section 3.2, we compare the action of the linear sigma model with the one of the D6-brane. 
We redefine the fields as eqs. (\ref{eq:redefinition1}), (\ref{eq:redefinition2}). To obtain them,
we look at all possible field redefinitions of the linear sigma model which make the same form 
of the action as the one of the D6-brane. 
It turns out that 
the field redefinition has an ambiguity and cannot be uniquely determined by just looking at the
terms in the cubic/quadratic Lagrangian terms. 
So, we take the simplest redefinition, which is eqs. (\ref{eq:redefinition1}), (\ref{eq:redefinition2}).
They look
\begin{equation}
\label{eq:a}
\begin{split} 
R &=f_{\pi}+\delta R+\epsilon_{1}(\delta R)^2+\epsilon_{2}(\delta\theta)^2+\epsilon_{3}(\delta R^{'})^2 \ , \\
\theta &=\qquad\delta\theta+\epsilon_{4}\delta R\delta\theta \ , 
\end{split}
\end{equation}
where $\delta R$ and $\delta\theta$ are the fluctuations around the static solution, $(R ,\, \theta)=(f_{\pi} ,\, 0 )$, respectively, and $\delta R':=\frac{\partial }{\partial t}\delta R$.
With this, 
the Lagrangian of the linear sigma model takes the same form as the one of the D6-brane up to cubic order in fluctuations.

Our simplification among all allowed redefinitions is: (1) we allow only up to a single derivative acting on each
fluctuation field. (2) we throw away $(\delta \theta')^2$ term in the redefinition.
Note that this simplification, leading to eqs. (\ref{eq:redefinition1}) and (\ref{eq:redefinition2}), still allows
a consistent field redefinition from the linear sigma model to the D4/D6 model.
In the following, we shall explain these ambiguities.

\noindent
\underline{(1) Higher derivatives}

We will show that 
we can add arbitrary higher derivatives to the redefinition as long as the the highest order is even. 
%
%
%
For instance, if we add to the right hand side of eqs. \eqref{eq:a}
the following terms to $R$ and $\theta$, respectively, we can obtain the same form of the Lagrangian to cubic order in fluctuations: 
\begin{align}
\label{eq:b}
&\delta V=\epsilon _{5}f_{\pi}^2(\delta\theta')^2+\epsilon_{6}f_{\pi}^2\delta\theta\delta\theta''
+\epsilon_{7}\bigg[\delta R'(\delta R)^{(5)}+\delta R''(\delta R)^{(4)}- \frac{1}{2}\left\{(\delta R)^{(3)}\right\}^2 \nonumber \\
&\qquad\quad\qquad\qquad\quad-\frac{\kappa}{2}\delta R(\delta R)^{(4)}-\kappa\delta R'(\delta R)^{(3)}+\frac{3\kappa}{4}(\delta R'')^2-\frac{3\kappa^2}{4}\delta R\delta R''\bigg]\quad
\mbox{to} \ R \ , \nonumber \\
&\delta W=2\epsilon_{5}\delta R'\delta\theta' -\epsilon_{6}\delta R''\delta\theta
\quad \mbox{to} \ \theta \ ,
\end{align}

\noindent
where $\kappa:=\mu^2+3g_{4}f_{\pi}^2,\,(\delta R)^{(n)}:= \frac{\partial^n }{\partial t^n}\delta R$.
To confirm this, substitute these terms into eq. \eqref{eq:c} written below.
When $\delta V$ and $\delta W$ represent the quadratic fluctuations in the redefinition of the fields $R$ and $\theta$, respectively ($\delta V$ and $\delta W$ are the functions of $\delta R, \delta\theta$ and their derivatives), the contribution of these terms to the Lagrangian of cubic order in fluctuations is
\begin{equation}
\label{eq:c}
\delta\mathcal{L}=\delta R'\delta V'+f_{\pi}^2\delta\theta'\delta W'-\kappa\delta R\delta V+af_{\pi}\delta\theta\delta W \ .
\end{equation}

\noindent
So, the resultant action is written by
\begin{multline}\label{eq:x}
S=\int dt\bigg[\frac{1}{2}\left\{(\delta R'(t))^2+f_{\pi}^2(\delta\theta '(t))^2\right\}-\frac{1}{2}\kappa(\delta R)^2+\frac{1}{2}f_{\pi}a(\delta\theta)^2\\
+\left(\frac{a}{2}-\epsilon_{2}\kappa+\epsilon_{4}f_{\pi}a\right)\delta R(\delta\theta)^2
-(g_{4}f_{\pi}+\epsilon_{1}\kappa)(\delta R)^3+\left(2\epsilon_{1}-\epsilon_{3}\kappa-\frac{3\epsilon_{7}}{2}\kappa^3\right)\delta R(\delta R')^2\\
+\left\{2\epsilon_{2}+\epsilon_{4}f_{\pi}^2+2(\epsilon_{5}+\epsilon_{6})f_{\pi}a+\epsilon_{6}f_{\pi}^2\kappa\right\}\delta R'\delta\theta ' \delta\theta
+\left\{f_{\pi}+\epsilon_{4}f_{\pi}^2-(\epsilon_{5}-\epsilon_{6})f_{\pi}^2\kappa\right\}\delta R(\delta\theta')^2\\
+(\mbox{surface terms)}'\bigg] \ , \nonumber
\end{multline}

\noindent
where
\begin{multline}
\mbox{(surface terms)}=\frac{2}{3}\epsilon_{3}(\delta R')^3+2\epsilon_{5}f_{\pi}^2\delta R'(\delta\theta')^2\\
+\epsilon_{6}\left\{f_{\pi}^2(\delta R'\delta\theta\delta\theta''-\delta R''\delta\theta\delta\theta')-f_{\pi}a\delta R'(\delta\theta)^2-f_{\pi}^2\kappa\delta R\delta\theta'\delta\theta\right\}+\epsilon_{7}\bigg[(\delta R')^2\delta R^{(5)}\\
+\kappa\left(-\frac{3}{2}\delta R\delta R'\delta R^{(4)}+\frac{1}{2}\delta R\delta R''\delta R^{(3)}\right)+\kappa^2\left\{\frac{1}{2}(\delta R)^2\delta R^{(3)}-\frac{3}{4}\delta R\delta R'\delta R''\right\}+\frac{3\kappa^3}{4}(\delta R)^2\delta R'\bigg] \ .
\end{multline}

\noindent
Renormalizing $f_{\pi}\delta\theta\rightarrow\delta\theta$, we can make this action take the same form as eq. \eqref{eq:d6flucaction}.

Furthermore, this ambiguity can be generalized to arbitrary order in derivatives, as long as the highest order of the differential of $\delta R$ is even. 
Define the sum of quadratic higher derivatives, $\delta S_{2n}, \delta T_{2(n-1)}, \mbox{and}\, P_{n}$, $(n\geq 2)$ by
\begin{equation}\label{eq:d}
\delta S_{2n}= \displaystyle\sum_{k=0}^n a_k\delta R^{(k)}\delta R^{(2n-k)} \ ,
\end{equation}
where\footnote{In $n=2$ case, this equation should be omitted. Equations \eqref{eq:h}, \eqref{eq:i} follow the same rule. } 
\begin{equation}\label{eq:e}
a_{0}=0, \quad a_{1}\neq0, \quad a_{k}=(-1)^k a_{1} \quad(2\leq k \leq n-1),\quad a_{n}=(-1)^n \frac{a_{1}}{2} \ , 
\end{equation}
and
\begin{equation}\label{eq:f}
\delta T_{2n-2}=\displaystyle\sum_{k=0}^{n-1}b_k\delta R^{(k)}\delta R^{(2n-k-2)} \ ,
\end{equation}
\begin{equation}\label{eq:g}
P_{n}=\displaystyle\sum_{k=1}^{n}p_k\delta R\delta R^{(k-1)}\delta R^{(2n-k)}\ ,
\end{equation}

\noindent
where $\{p_{k}\} \,(1\leq k \leq n)$ and  $\{b_{k}\} \,(0 \leq k \leq n-1)$ are given by the simultaneous equation
\begin{align}\label{eq:h}
\begin{cases}
p_{1}&=-\kappa a_{0} \ , \\
2p_{1}+p_{2}&=-\kappa a_{1}+b_{0} \ , \\
p_{k}+p_{k+1}&=-\kappa a_{k} \quad (2 \leq k \leq n-1) \ , \\
p_{n}&=-\kappa a_{n} \ .
\end{cases}
\end{align}
\begin{align}\label{eq:i}
\begin{cases}
b_{k-1}+b_{k}&=p_{k+1} \quad(1\leq k \leq n-2) \ , \\
b_{n-2}+2b_{n-1}&=p_{n} \ .
\end{cases}
\end{align}

\noindent
Then, we obtain 
\begin{equation}\label{eq:j}
\delta R'\delta S'_{2n}-\kappa\delta R\delta S_{2n}+\delta R'\delta T'_{2(n-1)}=\frac{\partial}{\partial t}\{a_{1}(\delta R')^2\delta R^{(2n-1)}+P_{n}\} \ .
\end{equation}
Note that $\delta R'\delta S'_{2n}$ is the sum of the $2(n+1)$th derivatives, while $-\kappa\delta R\delta S_{2n}$ and $\delta R'\delta T'_{2(n-1)}$ are those of the $2n$th derivatives, and $\delta T_{2(n-1)}$ is uniquely determined once $\delta S_{2n}$ is given.
Almost all the $2(n+1)$th derivatives in $\delta R'\delta S'_{2n}$ cancel each other, and remaining terms become surface terms. Combined with such contributions of $\delta T_{2(n-1)}$ as $\delta R'\delta T'_{2(n-1)}$ , all the $2n$th derivatives in $-\kappa\delta R\delta S_{2n}$ become surface terms. In consequence, even if $\delta S_{2n}$ and $\delta T_{2(n-1)}$ are added to $R$, all the contributions of the $2(n+1)$th and $2n$th derivatives to the action vanish. The remaining contributions are only those of $-\alpha\delta R\delta T_{2(n-1)}$, which are the $2(n-1)$th derivatives.
Thus, repeating the same procedure, and adding the sum of derivatives from $2n$th-order to second-order in turn, we obtain 
\begin{multline}\label{eq:k}
\delta R'\delta S'_{2n}-\kappa\delta R\delta S_{2n}+\delta R'\delta T'_{2(n-1)}-\kappa\delta R\delta T_{2(n-1)}+\ldots \\
+\delta R'\delta U'_{4}-\kappa\delta R\delta U_{4}+\delta R'\delta V'_{2}-\kappa\delta R\delta V_{2} \\
=(\mbox{surface terms})'-\kappa\delta R\delta V_{2} \ .
\end{multline}
$-\kappa\delta R\delta V_{2}$ is generically expressed as
\begin{equation}\label{eq:l}
\begin{split}
-\kappa\delta R\delta V_{2}&=-\kappa\delta R \{c_{0}\delta R\delta R''+c_{1}(\delta R')^2 \} \\
&=\frac{\partial}{\partial t}\{-c_0 \kappa(\delta R)^2\delta R'\}+\kappa(2c_{0}-c_{1})\delta R(\delta R')^2 \\
&=(\mbox{surface terms})'+(\mbox{terms included in the Lagrangian of D6 brane}) \ .
\end{split}
\end{equation}

\noindent
Therefore, even if we redefine the fields $R$ and $\theta$ as 
\begin{align}\label{eq:y}
&R=f_{\pi}+\delta R+\epsilon_{1}(\delta R)^2+\epsilon_{2}(\delta\theta)^2+\epsilon_{3}(\delta R')^2+\epsilon_{5}f_{\pi}^2(\delta\theta')^2+\epsilon_{6}f_{\pi}^2\delta\theta\delta\theta'' \nonumber \\
&\qquad\qquad\qquad\qquad\qquad\qquad\qquad\quad+\epsilon_{8}\{\delta S_{2n}+\delta T_{2(n-1)}+\ldots+\delta U_{4}+\delta V_{2}\} \ , \nonumber \\
&\theta=\qquad\delta\theta+\epsilon_{4}\delta R\delta\theta+\epsilon_{5}\delta R'\delta\theta'-\epsilon_{6}\delta R''\delta\theta \ ,
\end{align}
the resultant action takes the same form as eq. \eqref{eq:x} except for the coefficient of the term $\delta R(\delta R')^2$.

In this manner, in the redefinition of the fields of the linear sigma model, we can add arbitrary higher derivatives which satisfy eqs. \eqref{eq:d}, \eqref{eq:e} to $R$, and obtain the same action as the one of the D6-brane, so far as we add appropriate derivatives which can reduce all the higher derivatives to surface terms; that is, the field redefinition of the linear sigma model has such an ambiguity.
Taking this ambiguity into account, it is natural that 
we restrict the number of differentiations of $\delta R$ and $\delta\theta$  to at most $1$ when we redefine the fields to quadratic order, and compare the Lagrangians to cubic order in fluctuations.\footnote{ 
This criterion may be equivalent to have a natural Hamilton formalism.}

\vspace{5mm}

\noindent
\underline{(2) the $(\delta \theta')^2$ term}

Using the requirement that the derivative should be at most first order acting on the fluctuation fields, 
we can restrict the form of the field redefinitions to
\begin{equation}\label{eq:m}
\begin{split}
R&=f_{\pi}+\delta R+\epsilon_1(\delta R)^2+\epsilon_2(\delta\theta)^2+\epsilon_3(\delta R')^2+\epsilon_5f_{\pi}^2(\delta\theta')^2 \ , \\
\theta&=\hspace{9mm}\delta\theta+\epsilon_4(\delta R)(\delta\theta)+2\epsilon_5(\delta R')(\delta\theta') \ .
\end{split}
\end{equation}
This still differs by the $\epsilon_5$ terms, compared to \eqref{eq:a}, which causes a problem.
Though we should find the value of eight parameters,  $\epsilon_{1},\epsilon_{2},\ldots \epsilon_{5}, \mu^2, g_{4}$, and $a$, we have only seven equations in comparison between the Lagrangians of both models. In order to determine all the values, we must eliminate one parameter. So far, we have redefined the fields to quadratic order, and written down the Lagrangian to cubic order in fluctuations. To consider this problem, let us redefine the fields to cubic order, and express the Lagrangian to quartic order. Then, we will find which parameter should be eliminated.  With this extension, we relax the restriction addressed above in the following way. We now assume  that the order of differential of $\delta R$ and $\delta\theta$ is allowed up to second in the redefinition of the fields of cubic order.
Suppose $\epsilon_{5}=0$ in eqs. \eqref{eq:m}, and add the following terms
\begin{align}\label{eq:n}
\begin{split}
&(-2\epsilon_1\epsilon_3+\epsilon_3^2\kappa)\delta R(\delta R')^2+2\epsilon_3^2(\delta R')^2\delta R''\quad \mbox{to}\  R \ , \\
&-\frac{2\epsilon_2\epsilon_3}{f_{\pi}^2}(\delta R')^2\delta\theta+\tau\delta\theta(\delta\theta')^2\quad \mbox{to}\  \theta \ ,
\end{split}
\end{align}
where $\tau$ is a new parameter which is not included in eqs. \eqref{eq:m}. Then, the Lagrangian of the linear sigma model takes the same form as that of the D6-brane to quartic order in fluctuations.
On the other hand, if $\epsilon_{5}\neq 0$, then we must add the following awkward terms in addition to  eqs. \eqref{eq:n}, and set one of the parameters except for $\epsilon_5$ equal to zero. 
\begin{align}\label{eq:o}
&2\epsilon_5^2f_{\pi}^2\delta R''(\delta\theta')^2+2\epsilon_5(\epsilon_3+\epsilon_5)f_{\pi}^2\delta R'\delta\theta'\delta\theta''-2\epsilon_5f_{\pi}(\epsilon_4f_{\pi}+\epsilon_5a)\delta R'\delta\theta'\delta\theta\nonumber \\
&\qquad\qquad\qquad\qquad\qquad\qquad\qquad\qquad\qquad+\epsilon_5f_{\pi}^2(\epsilon_3\kappa-2\epsilon_1)\delta R(\delta\theta')^2\quad \mbox{to}\ R \ , \nonumber \\
&2\epsilon_{5}^2(\delta R')^2\delta\theta''+2\epsilon_5(\epsilon_3+\epsilon_5)\delta R'\delta R''\delta\theta'-\frac{\epsilon_3\epsilon_5a}{f_{\pi}}(\delta R')^2\delta\theta+2\epsilon_5(\epsilon_5\kappa-\epsilon_4-\frac{2}{f_{\pi}})\delta R\delta R'\delta\theta' \nonumber \\
&\qquad\qquad\qquad\qquad\qquad\qquad\qquad\qquad\qquad+2\epsilon_{5}^2f_{\pi}^2(\delta\theta')^2\delta\theta''-2\epsilon_2\epsilon_5\delta\theta(\delta\theta')^2
\quad \mbox{to} \ \theta \ .
\end{align}
Actually, there are other choices of the combinations of fluctuations to add. We here show one of the simplest choices.
The conclusion is that we set $\epsilon_{5}=0$, and adopt eqs. \eqref{eq:a} in order to simplify the redefinition of the fields in cubic order as possible. In the linear sigma model, since the action restores the rotation symmetry at $a\rightarrow 0$ limit, this ansatz also agrees with such a requirement.

\end{document}